\documentclass[preprint, 3p]{elsarticle}

\usepackage{hyperref}

\makeatletter
\def\ps@pprintTitle{%
	\let\@oddhead\@empty
	\let\@evenhead\@empty
	\let\@oddfoot\@empty
	\let\@evenfoot\@oddfoot
}
\makeatother

\graphicspath{{figures/}}
\usepackage{caption}
\usepackage{subcaption}
\usepackage{graphicx} 
\usepackage{amsmath,amssymb,amsfonts,amsthm,upgreek}
\usepackage{csquotes}
\usepackage{mathtools}
\usepackage{textcomp}
\usepackage{booktabs}
\usepackage{bm}
\usepackage[section]{placeins}
\usepackage{multirow}
\usepackage{algorithm}
\usepackage{algpseudocode}%
\usepackage{etoolbox}
\usepackage{tikz}
\usepackage{enumitem}

\DeclarePairedDelimiter\abs{\lvert}{\rvert}%
\begin{document}
	
	\begin{frontmatter}
		
		\title{On off-line and on-line Bayesian filtering for uncertainty quantification of structural deterioration}
		
		
		\author[mymainaddress,mytertiaryaddress]{Antonios Kamariotis \corref{mycorrespondingauthor}}
		\cortext[mycorrespondingauthor]{Corresponding author}
		\ead{antonis.kamariotis@tum.de}
		\author[mymainaddress]{Luca Sardi}
		\ead{luca.sardi@tum.de}
		\author[mymainaddress]{Iason Papaioannou}
		\ead{iason.papaioannou@tum.de}
		\author[mysecondaryaddress,mytertiaryaddress]{Eleni Chatzi}
		\ead{chatzi@ibk.baug.ethz.ch}
		\author[mymainaddress]{Daniel Straub}
		\ead{straub@tum.de}

		\address[mymainaddress]{Engineering Risk Analysis Group, Technical University of Munich, Theresienstrasse 90, 80333 Munich, Germany}
		\address[mysecondaryaddress]{Institute of Structural Engineering, ETH Zurich, Stefano-Franscini-Platz 5, 8093 Zurich, Switzerland}
		\address[mytertiaryaddress]{Institute for Advanced Study, Technical University of Munich, Lichtenbergstrasse 2a, 85748 Garching, Germany}
		
		\begin{abstract}
Data-informed predictive maintenance planning largely relies on stochastic deterioration models. Monitoring information can be utilized to update sequentially the knowledge on time-invariant deterioration model parameters either within an off-line (batch) or an on-line (recursive) Bayesian framework. With a focus on the quantification of the full parameter uncertainty, we review, adapt and investigate selected Bayesian filters for parameter estimation: an on-line particle filter, an on-line iterated batch importance sampling filter, which performs Markov chain Monte Carlo (MCMC) move steps, and an off-line MCMC-based sequential Monte Carlo filter. A Gaussian mixture model is used to approximate the posterior distribution within the resampling process in all three filters. Two numerical examples serve as the basis for a comparative assessment of off-line and on-line Bayesian estimation of time-invariant deterioration model parameters. The first case study considers a low-dimensional, nonlinear, non-Gaussian probabilistic fatigue crack growth model that is updated with sequential crack monitoring measurements. The second high-dimensional, linear, Gaussian case study employs a random field to model corrosion deterioration across a beam, which is updated with sequential measurements from sensors. The numerical investigations provide insights into the performance of off-line and on-line filters in terms of the accuracy of posterior estimates and the computational cost, when applied to problems of different nature, increasing dimensionality and varying sensor information amount. Importantly, they show that a tailored implementation of the on-line particle filter proves competitive with the computationally demanding MCMC-based filters. Suggestions on the choice of the appropriate method in function of problem characteristics are provided.
		\end{abstract}
		
		
	\end{frontmatter}
	
	
	\section*{Impact Statement}
Stochastic models describing time-evolving processes are widespread in science and engineering. In the modern data-rich engineering landscape, Bayesian methods can exploit monitoring data to sequentially update knowledge on underlying model parameters. The quantification of the full posterior uncertainty of these parameters is indispensable for several real-world tasks, where decisions need to be taken in view of the evaluated margins of risk and uncertainty. This work contributes to these tasks by rigorously reviewing the off-line and on-line Bayesian framework for the purpose of parameter estimation. On-line and off-line Bayesian filters are adapted and compared on a set of numerical examples of varying complexity related to structural deterioration. This results in suggestions regarding the suitability of each algorithm to specific applications.

\section{Introduction}
\label{sec:introduction}

Structural deterioration of various forms is present in most mechanical and civil structures and infrastructure systems. Accurate and effective tracking of structural deterioration processes can help to effectively manage it and minimize the total life-cycle costs \citep{Cadini_2009_b, Frangopol_2011, Kim_2017, Kamariotis_2022a}. The deployment of sensors on structural components/systems can enable  long-term monitoring of such processes. Monitoring data obtained sequentially at different points in time must be utilized in an efficient manner within a Bayesian framework to enable data-informed estimation and prediction of the deterioration process evolution.

Monitored structural deterioration processes are commonly modeled using Markovian state-space representations \citep{Myotyri_2006, Cadini_2009, Baraldi_2013}, whereby the deterioration state evolution is represented by a recursive Markov process equation, and is subject to stochastic process noise \citep{Corbetta_2018}. Monitoring information is incorporated by means of the measurement equation. The deterioration models further contain time-invariant uncertain parameters. The state-space can be augmented to include these parameters, if one wishes to obtain updated estimates thereof conditional on the monitoring information \citep{Straub_2009, Bhaskar_2009, Sun_2014, Corbetta_2018, Sang-ri_2018, Cristiani_2021, Kamariotis_2022b}; this is referred to as joint state-parameter estimation \citep{Sarkka_2013, Kantas_2015}. 

The formulation of a Markovian state-space representation of the deterioration process is not strictly required. The uncertain structural deterioration state is often defined solely as a function of uncertain time-invariant model parameters \citep{Ditlevsen_Madsen_1996, Vu_2000, Elingwood_2005, Stewart_2007}, which can be updated in view of the monitoring data. This updating, referred to herein as Bayesian parameter estimation, is often the primary task of interest. In this case, the deterioration state variables are obtained as outputs of the calibrated deterioration model with posterior parameter estimates \citep{Kennedy_2001, Ramancha_2022}. In spite of this, the problem of parameter estimation only can still be cast into a Markovian state-space representation. Quantifying the full posterior uncertainty of the time-invariant model parameters is essential for performing monitoring-informed predictions on the deterioration process evolution, the subsequent monitoring-informed estimation of the time-variant structural reliability \citep{Straub_2020, Melchers_2017} or the remaining useful life \citep{Sun_2014, Kim_2017}, and eventually for predictive maintenance planning.

Bayesian parameter estimation is the main focus of this paper. In long-term deterioration monitoring settings, where data is obtained sequentially at different points in time, Bayesian inference can be performed either in an on-line or an off-line framework \citep{Storvik_2002, Kantas_2015, Azam_2017}. In literature, these are also referred to as recursive (on-line) and batch (off-line) estimation \citep{Sarkka_2013}. Parameter estimation is cast into a state-space setup to render it suitable for application with on-line Bayesian filtering algorithms \citep{Kantas_2015}, such as the Kalman filter \citep{Kalman_1960} and its nonlinear variants \citep{Jazwinski_1970, Julier_1997, Daum_2005, SONG_2020}, the ensemble Kalman filter \citep{Evensen_2006}, and particle filters \citep{Doucet_2001,Doucet_2008,Sarkka_2013, Tatsis_2022}. We employ on-line particle filter methods for pure recursive estimation of time-invariant deterioration model parameters, which is not the typical use case for such methods, and can lead to degenerate and impoverished posterior estimates \citep{Del_Moral_2006,Sarkka_2013}. Taking that into account, we provide a formal investigation and discussion on the use of such methods for quantifying the full posterior uncertainty of time-invariant model parameters.

In its most typical setting within engineering applications, Bayesian parameter estimation is commonly performed with the use of off-line Markov Chain Monte Carlo (MCMC) methods, which have been used extensively in statistics and engineering to sample from complex posterior distributions of model parameters \citep{Hastings_1970, Gilks_1995, Beck_2002, Haario_2006, Ching_2007, Papaioannou_2015, Wu_2017, Lye_2021}. However, use of off-line methods for on-line estimation tasks is computationally prohibitive \citep{Del_Moral_2006, Kantas_2015}. Additionally, when considering off-line inference, in settings when measurements are obtained sequentially at different points in time, off-line MCMC methods tend to induce a larger computational cost than on-line particle filter methods, which can be important, e.g., when optimizing inspection and monitoring \citep{PAPAKONSTANTINOU_2014, Luque_2019, Kamariotis_2022b}. Questions that we investigate in this context include: Can one accurately quantify the uncertainty in the posterior parameter estimates when employing on-line particle filter methods for parameter estimation only purposes? How does this estimation compare against the posterior estimates obtained with off-line MCMC methods? How does the estimation accuracy depend on the nature of the problem, i.e., dimensionality, nonlinearity, or non-Gausssianity? What is the computational cost induced by the different methods? Ideally, one would opt for the method which can provide sufficiently accurate posterior results at the expense of the least computational cost. To address these questions, this paper reviews, adapts and selects algorithms in view of parameter estimation, and performs a comparative assessment of selected off-line and on-line filters specifically tailored for off-line and on-line Bayesian parameter estimation. The innovative comparative assessment results in a set of suggestions on the choice of the appropriate algorithm in function of problem characteristics.

The paper is structured as follows. Section \ref{sec:algorithms} provides a detailed description of on-line and off-line Bayesian inference in the context of parameter estimation only. Three different selected and adapted methods are presented in full algorithmic detail, namely an on-line particle filter with Gaussian mixture-based resampling (PFGM) \citep{Merwe_2003, McLachlan_2007}, the on-line iterated batch importance sampling filter (IBIS) \citep{Chopin_2002}, which performs off-line MCMC steps with a Gaussian mixture as a proposal distribution, and an off-line MCMC-based sequential Monte Carlo (SMC) filter \citep{Del_Moral_2006}, which enforces tempering of the likelihood function (known as simulated annealing) to sequentially arrive to the single final posterior density of interest \citep{Neal_2001, Jasra_2011}. The tPFGM and tIBIS variants, which adapt the PFGM and IBIS filters by employing tempering of the likelihood function of each new measurement, are further presented and proposed for problems with high sensor information amount. Section \ref{sec:Numerical_investigations} describes the two case studies that serve as the basis for numerical investigations, one non-linear, non-Gaussian and low-dimensional and one linear,  Gaussian and high-dimensional. MATLAB codes implementing the different algorithms and applying them on the two case studies introduced in this paper are made publicly available via a GitHub repository\footnote{\url{https://github.com/antoniskam/Offline_online_Bayes}}. Section 4 summarizes the findings of this comparative assessment, provides suggestions on choice of the appropriate method according to the nature of the problem, discusses cases which are not treated in our investigations, and concludes this work.

\section{On-line and off-line Bayesian filtering for time-invariant parameter estimation}
\label{sec:algorithms}

This work assumes the availability of a stochastic deterioration model $D$, parametrized by a vector $\boldsymbol{\theta}\in{\rm I\!R}^{d}$ containing the $d$ uncertain time-invariant model parameters. We collect the uncertain parameters influencing the deterioration process in the vector $\boldsymbol{\theta}$. In the Bayesian framework, $\boldsymbol{\theta}$ is modeled as a vector of random variables with a prior distribution $\pi_{\text{pr}}(\boldsymbol{\theta})$. We assume that the deterioration process is monitored via a permanently installed monitoring system. Long-term monitoring of a deterioration process leads to sets of noisy measurements $\{y_1,\dots, y_n\}$ obtained sequentially at different points in time $\{t_1,\dots, t_n\}$ throughout the lifetime of a structural component/system. Such measurements can be used to update the distribution of $\boldsymbol{\theta}$; this task is referred to as Bayesian parameter estimation. Within a deterioration monitoring setting, Bayesian parameter estimation can be performed either in an on-line or an off-line framework \citep{Kantas_2015}, depending on the task of interest. 

In an on-line framework, one is interested in updating the distribution of $\boldsymbol{\theta}$ in a sequential manner, i.e., at every time step $t_n$ when a new measurement $y_n$ becomes available, conditional on all measurements available up to $t_n$. Thus, in an on-line framework, inference of the sequence of posterior densities $\{\pi_{\text{pos}}(\boldsymbol{\theta}|\mathbf{y}_{1:n})\}_{n\geq 1}$ is the goal, where $\mathbf{y}_{1:n}$ denotes the components $\{y_1,\dots, y_n\}$. We point out that in this paper the term on-line does not relate to ``real-time" estimation, although on-line algorithms are also used in real-time estimation \citep{Chatzi_2009, Russel_2021}.

In contrast, in an off-line framework, inference of $\boldsymbol{\theta}$ is performed at a fixed time step $t_N$ using a fixed set of measurements $\{y_1,\dots,y_N\}$, and the single posterior density $\pi_{\text{pos}}(\boldsymbol{\theta}|\mathbf{y}_{1:N})$ is sought, which can be estimated via Bayes' rule as
\begin{equation}
	\pi_{\text{pos}}(\boldsymbol{\theta}|\mathbf{y}_{1:N}) \propto L(\mathbf{y}_{1:N}|\boldsymbol{\theta}) \pi_{\text{pr}}(\boldsymbol{\theta}),
	\label{eq:Bayes}
\end{equation}
where $L(\mathbf{y}_{1:N}|\boldsymbol{\theta})$ denotes the likelihood function of the whole measurement set $\mathbf{y}_{1:N}$ given the parameters $\boldsymbol{\theta}$. With the assumption that measurements are independent given the parameter state, $L(\mathbf{y}_{1:N}|\boldsymbol{\theta})$ can be expressed as a product of the likelihoods $L(y_n|\boldsymbol{\theta})$ as
\begin{equation}
	L(\mathbf{y}_{1:N}|\boldsymbol{\theta})=\prod_{n=1}^N L(y_n|\boldsymbol{\theta}).
	\label{eq:likelihood_product}
\end{equation}

MCMC methods sample from $\pi_{\text{pos}}(\boldsymbol{\theta}|\mathbf{y}_{1:N})$ via simulation of a Markov chain with $\pi_{\text{pos}}(\boldsymbol{\theta}|\mathbf{y}_{1:N})$ as its stationary distribution, e.g., by performing Metropolis Hastings (MH) steps \citep{Hastings_1970}. MCMC methods do not require estimation of the normalization constant in Equation \eqref{eq:Bayes}. However, in the on-line framework, MCMC methods are impractical, since they require simulating anew a different Markov chain for each new posterior $\pi_{\text{pos}}(\boldsymbol{\theta}|\mathbf{y}_{1:n})$, and the previously generated Markov chain for the posterior estimation of  $\pi_{\text{pos}}(\boldsymbol{\theta}|\mathbf{y}_{1:n-1})$ is not accounted for, except when choosing the seed for initializing the new Markov chain. This implies that MCMC methods quickly become computationally prohibitive in the on-line framework, already for a small $n$. An additional computational burden stems from the fact that each step within the MCMC sampling process requires evaluation of the full likelihood function $L(\mathbf{y}_{1:n}|\boldsymbol{\theta})$, i.e., the whole set of measurements $\mathbf{y}_{1:n}$ needs to be processed. This leads to increasing computational complexity for increasing $n$, and can render use of MCMC methods computationally inefficient even for off-line inference, especially when $N$ is large.

On-line particle filters \citep{Sarkka_2013, Kantas_2015} operate in a sequential fashion by making use of the Markovian property of the employed state-space representation, i.e., they compute $\pi_{\text{pos}}(\boldsymbol{\theta}|\mathbf{y}_{1:n})$ solely based on $\pi_{\text{pos}}(\boldsymbol{\theta}|\mathbf{y}_{1:n-1})$ and the new measurement $y_n$. The typical use of particle filters targets the tracking of a system's response (dynamic state) by means of a state-space representation \citep{Gordon_1993, Sarkka_2013}, while they are often also used also for joint state-parameter estimation tasks, wherein the state-space is augmented to include the model parameters to be estimated \citep{Sarkka_2013, Kantas_2015}. In addition, particle filters can also be applied for pure recursive estimation of time-invariant parameters, for which the noise in the dynamic model is formally zero \citep{Del_Moral_2006, Sarkka_2013}, although this is not the typical setting for application of particle filters. A model of the Markovian discrete time state-space representation for the case of time-invariant parameter estimation only is given in Equations \eqref{eq:state_space_a}, \eqref{eq:state_space_b}
\begin{subequations}
	\begin{align}
		\boldsymbol{\theta}_n &= \boldsymbol{\theta}_{n-1} \label{eq:state_space_a}\\
		y_n &= D_n\left(\boldsymbol{\theta}_n\right)\exp \left( \epsilon_n \right)\label{eq:state_space_b}
	\end{align}
	\label{eq:state_space}
\end{subequations}
where $\epsilon_n$ models the error/noise of the measurement at time $t_n$, and $\boldsymbol{\theta}_n$ denotes the time-invariant parameter vector at time step $n$. The dynamic equation for the time-invariant parameters \eqref{eq:state_space_a} is introduced for the sole purpose of casting the problem into a state-space representation. Since the measurements are assumed independent given the parameter state, the errors $\epsilon_n$ in Equation \eqref{eq:state_space_b} are independent. It should be noted that the measurement error, which is introduced in multiplicative form in Equation \eqref{eq:state_space_b}, is commonly expressed in an additive form \citep{Corbetta_2018}. Indeed, Equation \eqref{eq:state_space_b} can be reformulated in the logarithmic scale, whereby the measurement error is expressed in an additive form. All target distributions of interest in the sequence $\pi_{\text{pos}}(\boldsymbol{\theta}_n|\mathbf{y}_{1:n})$ are defined on the same space of $\boldsymbol{\theta}\in{\rm I\!R}^{d}$. In the remainder of this paper, the subscript $n$ will therefore be dropped from $\boldsymbol{\theta}_n$. As previously discussed, particle filters are mainly used for on-line inference. However, these can also be used in exactly the same way for off-line inference, where only a single posterior density $\pi_{\text{pos}}(\boldsymbol{\theta}|\mathbf{y}_{1:N})$ is of interest. In this case, particle filters use the sequence of measurements successively to sequentially arrive to the final single posterior density of interest via estimating all the intermediate distributions.

\subsection{On-line Particle Filter}
\label{subsec:PF}

Particle filter (PF) methods, also referred to as sequential Monte Carlo (SMC) methods, are importance sampling-based techniques that use a set of weighted samples $\{(\boldsymbol{\theta}_n^{(i)}, w_n^{(i)}):\ i=1,\dots, N_{\text{par}}\}$, called particles, to represent the posterior distribution of interest at estimation time step $n$, $\pi_{\text{pos}}(\boldsymbol{\theta}|\mathbf{y}_{1:n})$. PFs form the following approximation to the posterior distribution of interest:
\begin{equation}
	\pi_{\text{pos}}(\boldsymbol{\theta}|\mathbf{y}_{1:n}) \approx \sum_{i=1}^{N_{\text{par}}} w_n^{(i)} \delta (\boldsymbol{\theta}-\boldsymbol{\theta}_n^{(i)})
\end{equation}
where $\delta$ denotes the Dirac delta function.

\begin{figure}[!ht]
	\centering
	\includegraphics[width=0.50\textwidth]{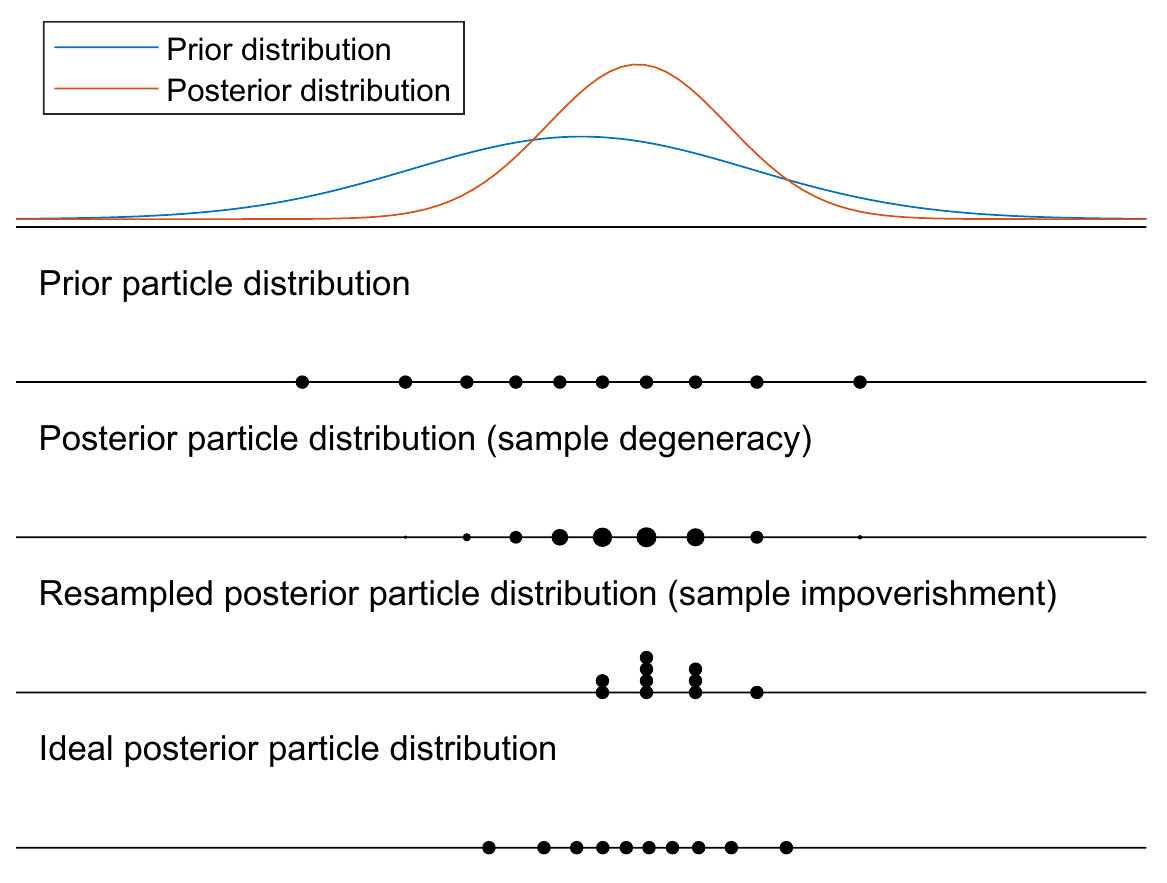}
	\caption{Sample degeneracy and impoverishment}
	\label{fig:degen_impov}
\end{figure}

When a new measurement $y_n$ becomes available, PFs shift from $\pi_{\text{pos}}(\boldsymbol{\theta}|\mathbf{y}_{1:n-1})$ to $\pi_{\text{pos}}(\boldsymbol{\theta}|\mathbf{y}_{1:n})$ by importance sampling using an appropriately chosen importance distribution, which results in a reweighting procedure (updating of the weights). An important issue that arises from this weight updating procedure is the sample degeneracy problem \citep{Sarkka_2013}. This relates to the fact that the importance weights $w_n^{(i)}$ become more unevenly distributed with each updating step. In most cases, after a certain number of updating steps, the weights of almost all the particles assume values close to zero (see Figure \ref{fig:degen_impov}). This problem is alleviated by the use of adaptive resampling procedures based on the effective sample size $N_{\text{eff}}=1/\sum_{i=1}^{N_{\text{par}}} \left(w_n^{(i)}\right)^2$ \citep{Liu_1998}. Most commonly, resampling is performed with replacement according to the particle weights whenever $N_{\text{eff}}$ drops below a user-defined threshold $N_{\text{T}} = cN_{\text{par}}, \ c \in[0,1]$. Resampling introduces additional variance to the parameter estimates \citep{Sarkka_2013}. In the version of the PF algorithm presented in Algorithm \ref{alg:pf}, the dynamic model of Equation \eqref{eq:state_space} is used as the importance distribution, as originally proposed in the bootstrap filter by \cite{Gordon_1993}.

\begin{algorithm}
	\caption{Particle Filter (PF)}\label{alg:pf}
	\begin{algorithmic}[1]%
		\vspace{0.1cm}
		\State generate $N_{\text{par}}$ initial particles $\boldsymbol{\theta}^{(i)}$ from $\pi_{\text{pr}}(\boldsymbol{\theta})$, \hspace{5mm} $i=1,\dots,N_{\text{par}}$
		\State assign initial weights $w_0^{(i)}=1/N_{\text{par}}$, \hspace{5mm} $i=1,\dots,N_{\text{par}}$
		\vspace{0.1cm}
		\For{$n=1,\dots,N$}
		\vspace{0.1cm}
		\State evaluate likelihood of the particles based on new measurement $y_n$, $L_n^{(i)} = L\left(y_n \mid \boldsymbol{\theta}^{(i)}\right) $
		\State update particle weights $w_n^{(i)} \propto L_n^{(i)} \cdot w_{n-1}^{(i)}$ and normalize $\mathrm{s.t.} \; \sum_{i=1}^{N_{\text{par}}} w_n^{(i)} = 1$
		\State evaluate $N_{\text{eff}}=\frac{1}{\sum_{i=1}^{N_{\text{par}}} \left(w_n^{(i)}\right)^2}$
		\If{$N_{\text{eff}}<N_{\text{T}}$}
		\State resample particles $\boldsymbol{\theta}^{(i)}$ with replacement according to $w_n^{(i)}$
		\State reset particle weights to $w_n^{(i)}=1/N_{\text{par}}$
		\EndIf
		\EndFor
	\end{algorithmic}
\end{algorithm}

When using PFs to estimate time-invariant parameters, for which the process noise in the dynamic equation is zero, one runs into the issue of sample impoverishment \citep{Sarkka_2013}. The origin of this issue is the resampling process. More specifically, after a few resampling steps, most (or in extreme cases all) of the particles in the sample set end up assuming the exact same value, i.e., the particle set consists of only few (or one) distinct particles (see Figure \ref{fig:degen_impov}). The sample impoverishment issue poses the greatest obstacle for time-invariant parameter estimation with PFs. A multitude of techniques have been suggested in literature to alleviate the sample impoverishment issue in joint state-parameter estimation setups \citep[see, e.g.,][]{Gilks_2001, Liu_2001, Musso_2001, Storvik_2002, Andrieu_2004, Andrieu_2010,  Chopin_2013, Chatzi_2013}. Fewer works have proposed solutions for resolving this issue in parameter estimation only setups \citep[see, e.g.,][]{Chopin_2002, Del_Moral_2006}. One of the simplest and most commonly used approaches consists of introducing artificial dynamics in the dynamic model of the parameter vector, i.e., the dynamic model $\boldsymbol{\theta}_n=\boldsymbol{\theta}_{n-1}+\boldsymbol{\epsilon}_{n-1}$ is employed, where $\boldsymbol{\epsilon}_{n-1}$ is a small artificial process noise \citep{Kitagawa_1998}. In this way, the time-invariant parameter vector is transformed into a time-variant one, therefore, the parameter estimation problem deviates from the original one \citep{Sarkka_2013, Kantas_2015}. This approach can introduce a bias and an artificial variance inflation in the estimates \citep{Kantas_2015}. For these reasons, this approach is not considered in this paper. 

To resolve the sample impoverishment issue encountered when using the PF Algorithm \ref{alg:pf} for parameter estimation only, this work employs the particle filter with Gaussian mixture resampling (PFGM), described in Algorithm \ref{alg:PF_GM}. The PFGM algorithm relates to pre-existing concepts \citep{Merwe_2003,Veettil_2016}, and is here specifically suggested for the parameter estimation only task, with its main goal being, in contrast to previous works, the quantification of the full posterior parameter uncertainty. A comparison between Algorithms \ref{alg:pf} and \ref{alg:PF_GM} shows that the only difference lies in the way that the resampling step is performed. PFGM replaces the standard resampling process of PF by first approximating the posterior distribution at estimation step $n$ by a Gaussian mixture model (GMM), which is fitted via the Expectation-Maximization (EM) algorithm \citep{McLachlan_2007, Chen_2010} on the weighted particle set. The degenerating particle set is then rejuvenated by sampling $N_{\text{par}}$ new particles from the GMM of Equation \eqref{eq:GMM},
\begin{equation}
	p\left( \boldsymbol{\theta} \mid \mathbf{y}_{1:n}\right) \approx \sum_{i=1}^{N_{\text{GM}}} \phi_{i} \mathcal{N} \left( \boldsymbol{\theta}; \boldsymbol{\mu}_{\mathbf{i}}, \mathbf{\Sigma_{i}} \right)
	\label{eq:GMM}
\end{equation}
where $\phi_{i}$ represents the weight of the Gaussian component $i$, while $\boldsymbol{\mu}_{\mathbf{i}}$ and $\mathbf{\Sigma_{i}}$ are the respective mean vector and covariance matrix. The number of Gaussians in the mixture $N_{\text{GM}}$, has to be chosen in advance, or can be estimated by use of appropriate algorithms \citep{Schubert_2017, Celeux_2019, Geyer_2019}. In the numerical investigations of Section \ref{sec:Numerical_investigations}, we set $N_{\text{GM}}$=8. We point out that the efficacy of PFGM strongly depends on the quality of the GMM posterior approximation. The reason for applying a GMM (and not a single Gaussian) is that the posterior distribution can deviate from the normal distribution, and can even be multimodal or heavy-tailed.

\begin{algorithm}
	\caption{Particle Filter with Gaussian mixture resampling (PFGM)}\label{pfgm}
	\begin{algorithmic}[1]%
		\vspace{0.1cm}
		\State generate $N_{\text{par}}$ initial particles $\boldsymbol{\theta}^{(i)}$ from $\pi_{\text{pr}}(\boldsymbol{\theta})$, \hspace{5mm} $i=1,\dots,N_{\text{par}}$
		\State assign initial weights $w_0^{(i)}=1/N_{\text{par}}$, \hspace{5mm} $i=1,\dots,N_{\text{par}}$
		\vspace{0.1cm}
		\For{$n=1,\dots,N$}
		\vspace{0.1cm}
		\State evaluate likelihood of the particles based on new measurement $y_n$, $L_n^{(i)} = L\left( y_n \mid \boldsymbol{\theta}^{(i)}\right) $
		\State update particle weights $w_n^{(i)} \propto L_n^{(i)} \cdot w_{n-1}^{(i)}$ and normalize $\mathrm{s.t.} \; \sum_{i=1}^{N_{\text{par}}} w_n^{(i)} = 1$
		\State evaluate $N_{\text{eff}}=\frac{1}{\sum_{i=1}^{N_{\text{par}}} \left(w_n^{(i)}\right)^2}$
		\If{$N_{\text{eff}}<N_{\text{T}}$}
		\State EM: fit a Gaussian mixture proposal distribution $g_{\text{GM}}(\boldsymbol{\theta})$ according to $\{\boldsymbol{\theta}^{(i)},w_n^{(i)}\}$
		\State sample $N_{\text{par}}$ new particles $\boldsymbol{\theta}^{(i)}$ from $g_{\text{GM}}(\boldsymbol{\theta})$ 
		\State reset particle weights to $w_n^{(i)}=1/N_{\text{par}}$
		\EndIf
		\EndFor
	\end{algorithmic}
	\label{alg:PF_GM}
\end{algorithm}

The simple reweighting procedure used in the on-line PFs is based on the premise that $\pi_{\text{pos}}(\boldsymbol{\theta}|\mathbf{y}_{1:n-1})$ and $\pi_{\text{pos}}(\boldsymbol{\theta}|\mathbf{y}_{1:n})$ are likely to be similar, i.e., that the new measurement $y_n$ will not cause a very large change in the posterior. However, when that is not the case, this simple reweighting procedure is bound to perform poorly, leading to very fast degeneration of the particle set. In cases where already the first measurement set $y_1$ is strongly informative relative to the prior, the PF is bound to strongly degenerate already in the first weight updating step (e.g., we observe this in the second case study of Section \ref{subsec:RF} in the case of 10 sensors). To counteract this issue, in this paper we incorporate the idea of simulated annealing (enforcing tempering of the likelihood function) \citep{Neal_2001} when needed within the on-line PFGM algorithm, which we term the tPFGM Algorithm \ref{alg:pf_gm_temp}. The tPFGM algorithm draws inspiration from previous works \citep{Gall_2007, Deutscher_2000}, but is here suggested for the parameter estimation only task, opting for the quantification of the full posterior parameter uncertainty. The algorithm operates as follows: At estimation time step $n$, before performing the reweighting operation, the algorithm first checks the updated effective sample size for indication of sample degeneracy. If no degeneracy is detected, tPFGM operates exactly like PFGM. When sample degeneracy occurs, tPFGM employs adaptive tempering of the likelihood $L\left(y_n \mid \boldsymbol{\theta}\right)$ of the new measurement $y_n$ in order to ``sequentially" sample from $\pi_{\text{pos}}(\boldsymbol{\theta}|\mathbf{y}_{1:n-1})$ to $\pi_{\text{pos}}(\boldsymbol{\theta}|\mathbf{y}_{1:n})$ by visiting a sequence of artificial intermediate posteriors, as defined by the tempered likelihood function $L^q\left(y_n \mid \boldsymbol{\theta}\right)$. The tempering factor $q$ takes values between 0 and 1. When $q=0$, the new measurement $y_n$ is neglected, while $q=1$ entails considering the whole likelihood function of $y_n$, thus reaching to $\pi_{\text{pos}}(\boldsymbol{\theta}|\mathbf{y}_{1:n})$. The intermediate values of $q$ are adaptively selected via solution of the optimization problem in line 11 of Algorithm \ref{alg:pf_gm_temp}, which ensures that the effective sample size does not drop below the threshold $N_{\text{T}}$ for the chosen $q$ value. Naturally, use of tPFGM can trigger more resampling events than PFGM, as resampling can occur more than once within a time step $n$.

\begin{algorithm}[!ht]
	\caption{Particle Filter with Gaussian mixture resampling and likelihood tempering (tPFGM)}\label{alg:pf_gm_temp}
	\begin{algorithmic}[1]%
		\vspace{0.1cm}
		\State generate $N_{\text{par}}$ initial particles $\boldsymbol{\theta}^{(i)}$ from $\pi_{\text{pr}}(\boldsymbol{\theta})$, \hspace{5mm} $i=1,\dots,N_{\text{par}}$
		\State assign initial weights $w_0^{(i)}=1/N_{\text{par}}$, \hspace{5mm} $i=1,\dots,N_{\text{par}}$
		\vspace{0.1cm}
		\For{$n=1,\dots,N$}
		\vspace{0.1cm}
		\State evaluate likelihood of the particles based on new measurement $y_n$, $L_n^{(i)} = L\left(y_n \mid \boldsymbol{\theta}^{(i)}\right) $
		\State set $q=0$ and create auxiliary particle weights $w_a^{(i)}=w_{n-1}^{(i)}$
		\While{$q\neq1$}
		\If {$N_{\text{eff}} = \left(\sum_{i=1}^{N_{\text{par}}} w_a^{(i)} \cdot {L_n^{(i)}}^{1-q} \right)^2 / \sum_{i=1}^{N_{\text{par}}} \left(w_a^{(i)} \cdot {L_n^{(i)}}^{1-q}\right)^2 >N_{\text{T}}$}
		\State  update auxiliary particle weights $w_a^{(i)} \propto w_a^{(i)}  \cdot {L_n^{(i)}}^{1-q}$ and normalize $\mathrm{s.t.} \sum_{i=1}^{N_{\text{par}}} w_a^{(i)} = 1$
		\State set $q=1$
		\Else
		\State solve $\left(\sum_{i=1}^{N_{\text{par}}} w_a^{(i)} \cdot {L_n^{(i)}}^{dq} \right)^2 / \sum_{i=1}^{N_{\text{par}}} \left(w_a^{(i)} \cdot {L_n^{(i)}}^{dq}\right)^2 - N_{\text{T}} = 0$ for $dq$
		\State set $q_{\text{new}} = \min \left[q+dq,1\right]$
		\State set $dq=q_{\text{new}}-q$ and $q=q_{\text{new}}$
		\State  update auxiliary particle weights $w_a^{(i)} \propto w_a^{(i)}  \cdot {L_n^{(i)}}^{dq}$ and normalize $\mathrm{s.t.} \sum_{i=1}^{N_{\text{par}}} w_a^{(i)} = 1$
		\State EM: fit a Gaussian mixture proposal distribution $g_{\text{GM}}(\boldsymbol{\theta})$ according to $\{\boldsymbol{\theta}^{(i)},w_a^{(i)}\}$
		\State sample $N_{\text{par}}$ new particles $\boldsymbol{\theta}^{(i)}$ from $g_{\text{GM}}(\boldsymbol{\theta})$ 
		\State reset auxiliary particle weights to $w_a^{(i)}=1/N_{\text{par}}$
		\EndIf						
		\EndWhile
		\State set $w_n^{(i)}=w_a^{(i)}$
		\EndFor
	\end{algorithmic}
\end{algorithm}

The PFGM and tPFGM filters rely entirely on the posterior approximation via a GMM for sampling $N_{\text{par}}$ new particles during the resampling process. However, there is no guarantee that these new particles follow the true posterior distribution of interest. The IBIS filter of the following Section \ref{subsec:IBIS} aims at addressing this issue. 

\subsection{Iterated Batch Importance Sampling}
\label{subsec:IBIS}

Implementing MCMC steps within PF methods to move the particles after a resampling step was originally proposed by \cite{Gilks_2001}, in the so-called resample-move algorithm. \cite{Chopin_2002} introduced a special case of the resample-move algorithm, specifically tailored for application to static parameter estimation only purposes, namely the iterated batch importance sampling (IBIS) filter. IBIS was originally introduced as an iterative method for solving off-line estimation tasks by incorporating the sequence of measurements one at a time. In doing this, the algorithm visits the sequence of intermediate posteriors within its process, and can therefore also be used to perform on-line estimation tasks. An on-line version of the IBIS filter is presented in Algorithm \ref{alg:ibis}, used in conjuction with the MCMC routine of Algorithm \ref{alg:mh_gm}.

The core idea of the IBIS filter is the following: At estimation step $n$, if sample degeneracy is identified, first the particles are resampled with replacement, and subsequently the resampled particles are moved with a Markov chain transition kernel whose stationary distribution is $\pi_{\text{pos}}(\boldsymbol{\theta}|\mathbf{y}_{1:n})$. More specifically, each of the $N_{\text{par}}$ resampled particles is used as the seed to perform a single MCMC step. This approach is inherently different to standard applications of MCMC, where a transition kernel is applied multiple times on one particle.

A question that arises is how to choose the Markov chain transition kernel. \cite{Chopin_2002} argues for choosing a transition kernel that ensures that the proposed particle only weakly depends on the seed particle value. It is therefore recommended to use an independent Metropolis-Hastings (IMH) kernel, wherein the proposed particle is sampled from a proposal distribution $g$, which has to be as close as possible to the target distribution $\pi_{\text{pos}}(\boldsymbol{\theta}|\mathbf{y}_{1:n})$. In obtaining such a proposal distribution, along the lines of what is described in Section \ref{subsec:PF}, in this work we employ a GMM approximation (see Equation \eqref{eq:GMM}) of the target distribution as the proposal density $g_{\text{GM}}(\boldsymbol{\theta})$ within the IMH kernel \citep{Papaioannou_2016, South_2019}. The IMH kernel with a GMM proposal distribution is denoted IMH-GM herein. The acceptance probability (line 6 of Algorithm \ref{alg:mh_gm}) of the IMH-GM kernel is a function of both the initial seed particle and the GMM proposed particle. The acceptance rate can indicate how efficient the IMH-GM kernel is in performing the MCMC move step within the IBIS algorithm. It is important to note that when computing the acceptance probability, a call of the full likelihood function is invoked, which requires the whole set of measurements $y_{1:n}$ to be processed; this leads to a significant additional computational demand, which pure on-line methods are not supposed to accommodate \citep{Doucet_2001}.

The performance of the IBIS sampler depends highly on the mixing properties of the IMH-GM kernel. If the kernel leads to slowly decreasing chain auto-correlation, the moved particles are bound to remain in regions close to the particles obtained by the resampling step. This can lead to an underrepresentation of the parameter space of the intermediate posterior distribution. It might therefore be beneficial to add a burn-in period within the IMH-GM kernel \citep{Del_Moral_2006}. Implementing that is straightforward and is shown in Algorithm \ref{alg:mh_gm}, where $n_\text{B}$ is the user-defined number of burn-in steps. Naturally, the computational cost of the IMH-GM routine increases linearly with the number of burn-in steps.

\begin{algorithm}
	\caption{Independent Metropolis Hastings with GM proposal (IMH-GM)}\label{alg:mh_gm}
	\begin{algorithmic}[1]%
		\vspace{0.1cm}
		\State \textbf{IMH-GM Input}:  $\{\boldsymbol{\theta}^{(i)},L^{(i)}\cdot \pi_{\text{pr}}(\boldsymbol{\theta}^{(i)})\}$, $\pi_{\text{pr}}(\boldsymbol{\theta})$, $L(\mathbf{y}_{1:n} | \boldsymbol{\theta})$ and $g_{\text{GM}}(\boldsymbol{\theta})$
		\For{$i=1,\dots,N_{\text{par}}$}
		\vspace{0.1cm}
		\For{$j=1,\dots,n_\text{B}+1$}
		\vspace{0.1cm}
		\State sample candidate particle $\boldsymbol{\theta}^{(i)}_{c,j}$ from $g_{\text{GM}}(\boldsymbol{\theta})$
		\State evaluate $L^{(i)}_{c,j}= L(\mathbf{y}_{1:n} | \boldsymbol{\theta}^{(i)}_{c,j}) $ for candidate particle
		\State evaluate acceptance ratio $\alpha = \min\left[1,\frac{L^{(i)}_{c,j}\cdot \pi_{\text{pr}}(\boldsymbol{\theta}^{(i)}_{c,j}) \cdot g_{\text{GM}}(\boldsymbol{\theta}^{(i)})}{L^{(i)}\cdot \pi_{\text{pr}}(\boldsymbol{\theta}^{(i)}) \cdot g_{\text{GM}}(\boldsymbol{\theta}^{(i)}_{c,j})}\right]$
		\State generate uniform random number $u \in [0,1]$
		\If{$u<\alpha$}
		\State replace $\{\boldsymbol{\theta}^{(i)},L^{(i)}\cdot \pi_{\text{pr}}(\boldsymbol{\theta}^{(i)})$\} with $\{\boldsymbol{\theta}^{(i)}_{c,j},L^{(i)}_{c,j}\cdot \pi_{\text{pr}}(\boldsymbol{\theta}^{(i)}_{c,j})$\}
		\EndIf
		\EndFor
		\EndFor
		\State \textbf{IMH-GM Output}: $\{\boldsymbol{\theta}^{(i)},L^{(i)}\cdot \pi_{\text{pr}}(\boldsymbol{\theta}^{(i)})\}$
	\end{algorithmic}
\end{algorithm}

Algorithm \ref{alg:ibis} details the workings of the IMH-GM-based IBIS filter used in this work. In line 11 of this algorithm, the IMH-GM routine of Algorithm \ref{alg:mh_gm} is called, which implements the IMH-GM kernel for the MCMC move step. Comparing Algorithms \ref{alg:PF_GM} and \ref{alg:ibis}, it is clear that both filters can be used for on-line inference within a single run, but the IBIS filter has significantly larger computational cost, as will also be demonstrated in the numerical investigations of Section \ref{sec:Numerical_investigations}. In the same spirit as the proposed tPFGM algorithm \ref{alg:pf_gm_temp}, which enforces simulated annealing (tempering of the likelihood function) in cases when $\pi_{\text{pos}}(\boldsymbol{\theta}|\mathbf{y}_{1:n-1})$ and $\pi_{\text{pos}}(\boldsymbol{\theta}|\mathbf{y}_{1:n})$ are likely to be quite different, the same idea can be implemented also within the IBIS algorithm. That leads to what we refer to as the tIBIS algorithm in this paper.

\begin{algorithm}
	\caption{IMH-GM-based Iterated Batch Importance Sampling (IBIS)}\label{alg:ibis}
	\begin{algorithmic}[1]%
		\vspace{0.1cm}
		\State generate $N_{\text{par}}$ initial particles $\boldsymbol{\theta}^{(i)}$ from $\pi_{\text{pr}}(\boldsymbol{\theta})$, \hspace{5mm} $i=1,\dots,N_{\text{par}}$
		\State assign initial weights $w_0^{(i)}=1/N_{\text{par}}$, \hspace{5mm} $i=1,\dots,N_{\text{par}}$
		\vspace{0.1cm}
		\For{$n=1,\dots,N$}
		\vspace{0.1cm}
		\State evaluate likelihood of the particles based on new measurement $y_n$, $L_n^{(i)} = L\left( y_n \mid \boldsymbol{\theta}^{(i)}\right) $
		\State evaluate the new target distribution, $L(\mathbf{y}_{1:n} | \boldsymbol{\theta}^{(i)}) \cdot \pi_{\text{pr}}\left(\boldsymbol{\theta}^{(i)}\right) = L_n^{(i)} \cdot L(\mathbf{y}_{1:n-1} | \boldsymbol{\theta}^{(i)}) \cdot \pi_{\text{pr}}\left(\boldsymbol{\theta}^{(i)}\right)$
		\State update particle weights $w_n^{(i)} \propto L_n^{(i)} \cdot w_{n-1}^{(i)}$ and normalize $\mathrm{s.t.} \; \sum_{i=1}^{N_{\text{par}}} w_n^{(i)} = 1$
		\State evaluate $N_{\text{eff}}=\frac{1}{\sum_{i=1}^{N_{\text{par}}} \left(w_n^{(i)}\right)^2}$
		\If{$N_{\text{eff}}<N_{\text{T}}$}
		\State EM: fit a Gaussian mixture proposal distribution $g_{\text{GM}}(\boldsymbol{\theta})$ according to $\{\boldsymbol{\theta}^{(i)},w_n^{(i)}\}$
		\State resample $N_{\text{par}}$ new particles $\{\boldsymbol{\theta}^{(i)},L(\mathbf{y}_{1:n} | \boldsymbol{\theta}^{(i)}) \cdot \pi_{\text{pr}}(\boldsymbol{\theta}^{(i)})\}$ with replacement according to $w_n^{(i)}$
		\State IMH-GM step with inputs $\{\boldsymbol{\theta}^{(i)},L(\mathbf{y}_{1:n} | \boldsymbol{\theta}^{(i)}) \cdot \pi_{\text{pr}}(\boldsymbol{\theta}^{(i)})\}$,  $\pi_{\text{pr}}(\boldsymbol{\theta})$, $L(\mathbf{y}_{1:n} | \boldsymbol{\theta})$ and $g_{\text{GM}}(\boldsymbol{\theta})$
		\State reset particle weights to $w_n^{(i)}=1/N_{\text{par}}$
		\EndIf
		\EndFor
	\end{algorithmic}
\end{algorithm}

\subsection{Off-line Sequential Monte Carlo sampler}
\label{subsec:SMC}
In Section 4 of \cite{Del_Moral_2006}, the authors presented a generic approach to convert an off-line MCMC sampler into a sequential Monte Carlo (SMC) sampler tailored for performing off-line estimation tasks, i.e., for estimating the single posterior density of interest $\pi_{\text{pos}}(\boldsymbol{\theta}|\mathbf{y}_{1:N})$. The off-line SMC sampler used in this work is presented in Algorithm \ref{smc} based on \cite{Del_Moral_2006} and \cite{Jasra_2011}. The key idea of this sampler is to adaptively construct the following artificial sequence of densities,
\begin{equation}
	\pi_{j}(\boldsymbol{\theta}|\mathbf{y}_{1:N}) \propto L^{q_j}(\mathbf{y}_{1:N}|\boldsymbol{\theta}) \pi_{\text{pr}}(\boldsymbol{\theta})
	\label{eq:Bayes_temp}
\end{equation}
where $q_j$ is a tempering parameter which obtains values between 0 and 1, in order to ``sequentially" sample in a smooth manner from the prior to the final single posterior density of interest. Once $q_j = 1$, $\pi_{\text{pos}}(\boldsymbol{\theta}|\mathbf{y}_{1:N})$ is reached. Similar to what was described in tPFGM, the intermediate values of $q_j$ are adaptively found via solution of the optimization problem in line 5 of Algorithm \ref{smc}. The GMM approximation of the intermediate posteriors and the IMH-GM kernel of Algorithm \ref{alg:mh_gm} in order to move the particles after resampling are also key ingredients of this SMC sampler. Unlike PFGM and IBIS, this SMC algorithm cannot provide the on-line solution within a single run, and has to be rerun from scratch for every new target posterior of interest. In this regard, use of Algorithm \ref{smc} for on-line inference is impractical.

\begin{algorithm}[!ht]
	\caption{IMH-GM-based Sequential Monte Carlo (SMC)}\label{smc}
	\begin{algorithmic}[1]%
		\vspace{0.1cm}
		\State generate $N_{\text{par}}$ initial particles $\boldsymbol{\theta}^{(i)}$ from $\pi_{\text{pr}}(\boldsymbol{\theta})$, \hspace{5mm} $i=1,\dots,N_{\text{par}}$
		\State evaluate for every particle the full likelihood $L^{(i)} = L(\mathbf{y}_{1:N}\mid\boldsymbol{\theta}^{(i)})$ and the prior $\pi_{\text{pr}}(\boldsymbol{\theta}^{(i)})$
		\State set $q=0$
		\While{$q\neq1$}
		\State solve $\left(\sum_{i=1}^{N_{\text{par}}} {L^{(i)}}^{dq}\right)^2 / \sum_{i=1}^{N_{\text{par}}} {L^{(i)}}^{2\cdot dq} - N_{\text{T}} = 0$ for $dq$
		\State set $q_{\text{new}} = \min \left[q+dq,1\right]$
		\State set $dq=q_{\text{new}}-q$ and $q=q_{\text{new}}$
		\State evaluate particle weights $w^{(i)} \propto {L^{(i)}}^{dq}$ and normalize $\mathrm{s.t.} \; \sum_{i=1}^{N_{\text{par}}} w^{(i)} = 1$
		\State EM: fit a Gaussian mixture proposal distribution $g_{\text{GM}}(\boldsymbol{\theta})$ according to $\{\boldsymbol{\theta}^{(i)},w^{(i)}\}$
		\State resample $N_{\text{par}}$ new particles $\{\boldsymbol{\theta}^{(i)},{L^{(i)}}^{q}\cdot \pi_{\text{pr}}(\boldsymbol{\theta}^{(i)})\}$ with replacement according to $w^{(i)}$
		\State IMH-GM step with inputs $\{\boldsymbol{\theta}^{(i)}, {L^{(i)}}^{q} \cdot \pi_{\text{pr}}(\boldsymbol{\theta}^{(i)})\}$ , $\pi_{\text{pr}}(\boldsymbol{\theta})$, $L^q(\mathbf{y}_{1:N} | \boldsymbol{\theta})$ and $g_{\text{GM}}(\boldsymbol{\theta})$
		\State reset particle weights to $w^{(i)}=1/N_{\text{par}}$
		\EndWhile
	\end{algorithmic}
\end{algorithm}

\subsection{Computational remarks}
The algebraic operations in all presented algorithms are implemented in the logarithmic scale, which employs evaluations of the logarithm of the likelihood function and, hence, ensures computational stability. Furthermore, the EM step for fitting the GMM is performed after initially transforming the prior joint probability density function of $\boldsymbol{\theta}$ to an underlying vector $\boldsymbol{u}$ of independent standard normal random variables \citep{Der_Kiureghian_1986}. In standard normal space, the parameters are decorrelated, which enhances the performance of the EM algorithm.

\section{Numerical investigations}
\label{sec:Numerical_investigations}

\subsection{Low-dimensional case study: Paris-Erdogan fatigue crack growth model}
\label{subsec:CGM}

A fracture mechanics-based model serves as the first case study. This describes the fatigue crack growth evolution under increasing stress cycles \citep{Paris_1963, Ditlevsen_Madsen_1996}. The crack growth follows the following first-order differential Equation \eqref{eq:ODE}, known as Paris-Erdogan law,
\begin{equation}
	\frac{da\left( n\right) }{dn} = \exp \left( C_{\ln} \right) \left[\Delta S \sqrt{\pi a \left( n \right)} \right]^m
	\label{eq:ODE}
\end{equation}
where $a \left[\text{mm}\right]$ is the crack length, $n \left[-\right]$ is the number of stress cycles, $\Delta S \left[\text{Nmm}^{-2}\right]$ is the stress range per cycle when assuming constant stress amplitudes, $C$ and $m$ represent empirically determined model parameters; $C_{\ln}$ corresponds to the natural logarithm of $C$.

The solution to this differential equation, with boundary condition $a\left(n=0\right)=a_0$, can be written as a function of the number of stress cycles $n$ and the vector $\boldsymbol{\theta}=\left[a_0, \Delta S, C_{ln}, m\right]$ containing the uncertain time-invariant model parameters as
\begin{equation}
	a\left( n, \boldsymbol{\theta} \right) = \left[ \left( 1- \frac{m}{2} \right) \exp \left( C_{\ln} \right) \Delta S^m \pi^{{m}\slash{2}} n +a_0^{\left(1-{m}\slash{2}\right)} \right]^{{\left(1-{m}\slash{2}\right)}^{-1}}
	\label{eq:ODE_solution}
\end{equation}

We assume that noisy measurements of the crack $y_n$ are obtained sequentially at different values of $n$. The measurement Equation \eqref{eq:meas_eq} assumes a multiplicative lognormal measurement error, $\exp \left( \epsilon_n \right)$.
\begin{equation}
	y_{n} = a_n(\boldsymbol{\theta}) \exp \left( \epsilon_n \right)
	\label{eq:meas_eq}
\end{equation}

Under this assumption, the likelihood function for a measurement at a given $n$ is shown in Equation \eqref{eq:likelihood}.
\begin{equation}
	L \big( y_{n} ; a_n\left(\boldsymbol{\theta}\right) \big) = \frac{1}{\sigma_{\epsilon_n} \sqrt{2 \pi}} \exp \left[ - \frac{1}{2} \left(\frac{\ln \left( y_n \right) - \mu_{\epsilon_n} - \ln \big( a_n\left(\boldsymbol{\theta}\right) \big)}{\sigma_{\epsilon_n}} \right)^2 \right]
	\label{eq:likelihood}
\end{equation}

Table \ref{tab:par_cgm} shows the prior probability distribution model for each random variable in the vector $\boldsymbol{\theta}$ \citep{Ditlevsen_Madsen_1996, Straub_2009}, as well as the assumed probabilistic model of the measurement error. In this case study we are dealing with a non-linear model and a parameter vector with non-Gaussian prior distribution.

\begin{table}[!ht]
	\caption{Prior distribution model for the fatigue crack growth model parameters and the measurement error}
	\small
	\begin{tabular*}{\textwidth}{@{\extracolsep{\fill}} c c c c c}
		\hline
		\textbf{Parameter} & \textbf{Distribution} & \textbf{Mean} & \textbf{Standard Deviation} & \textbf{Correlation}\\
		\hline
		$a_0$ & Exponential & $1$ & $1$ & $-$\\
		$\Delta S$ & Normal & $60$ & $10$ & $-$\\
		$C_{\ln}, \, m$ & Bi-Normal & $\left( -33; \, 3.5 \right)$ & $\left( 0.47; \, 0.3 \right)$ & $\rho_{C_{\ln},m} = -0.9$\\
		$\exp \left( \epsilon_n \right)$ & Log-normal & $1.0$ & $0.1508$& $-$\\
		\hline
	\end{tabular*}
	\label{tab:par_cgm}
\end{table}

\subsubsection{Markovian state-space representation for application of on-line filters}
\label{subsubsec:state_space_CGM}
A Markovian state-space representation of the deterioration process is required for application of on-line filters. The dynamic and measurement equations of the discrete-time state-space representation of the fatigue crack growth model with unknown time-invariant parameters $\boldsymbol{\theta}=\left[a_0, \Delta S, C_{ln}, m\right]$ are shown below.
\begin{equation}
	\begin{split}
		\boldsymbol{\theta}_k &= \boldsymbol{\theta}_{k-1} \\
		y_k &= a_k\left(\boldsymbol{\theta}_k\right)\exp \left( \epsilon_k \right) = \left[ \left( 1- \frac{m_k}{2} \right) \exp \left( C_{ln_k} \right) \Delta S_k^{m_k} \pi^{{m_k}\slash{2}} n +a_{0_k}^{\left(1-{m_k}\slash{2}\right)} \right]^{{\left(1-{m_k}\slash{2}\right)}^{-1}} \exp \left( \epsilon_k \right)
		\label{eq:state_space_CGM}
	\end{split}
\end{equation}

The subscript $k$ denotes the estimation time step. More specifically, the number of stress cycles is discretized as $n=k\Delta n$, with $k=1,\ldots,100$ and $\Delta n = 1\times10^5$. The state-space model of Equation \ref{eq:state_space_CGM} is nonlinear and the prior is non-Gaussian. For reasons explained in Section \ref{sec:algorithms}, the subscript $k$ in $\boldsymbol{\theta}_k$ is dropped in the remainder of this section.

\subsubsection{Reference posterior solution}
\label{subsubsec:reference_post_CGM}
\begin{figure}
	\centering
	\includegraphics[width=0.75\textwidth]{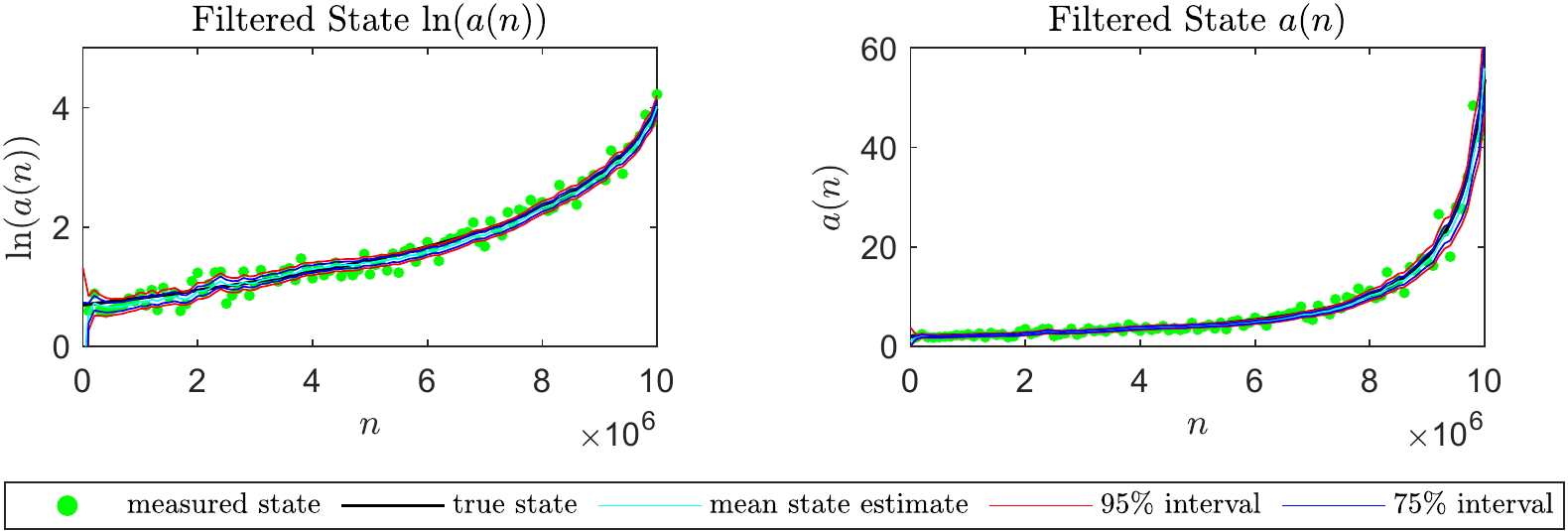}
	\caption{Reference mean and credible intervals for the filtered crack growth state $a_n$}
	\label{fig:cgm_ref_state_filt}
\end{figure}

For the purpose of performing a comparative assessment of the different filters, an underlying ``true" realization of the fatigue crack growth process $a^\ast(n)$ is generated for $n=k\Delta n$, with $k=1,\ldots,100$ and $\Delta n = 1\times10^5$. This realization corresponds to the randomly generated ``true" vector of time-invariant parameters $\boldsymbol{\theta}^\ast = [a_0^\ast=2.0, \Delta S^\ast=50.0, C_{ln}^\ast=-33.5, m^\ast=3.7]$. Sequential synthetic crack monitoring measurements $y_k$ are sampled from the measurement Equation \eqref{eq:meas_eq} for $a_k(\boldsymbol{\theta}^\ast)$, and for randomly generated measurement noise samples $\exp \left( \epsilon_k \right)$. These measurements are scattered in green in Figure \ref{fig:cgm_ref_state_filt}.

\begin{figure}
	\centering
	\includegraphics[width=0.85\textwidth]{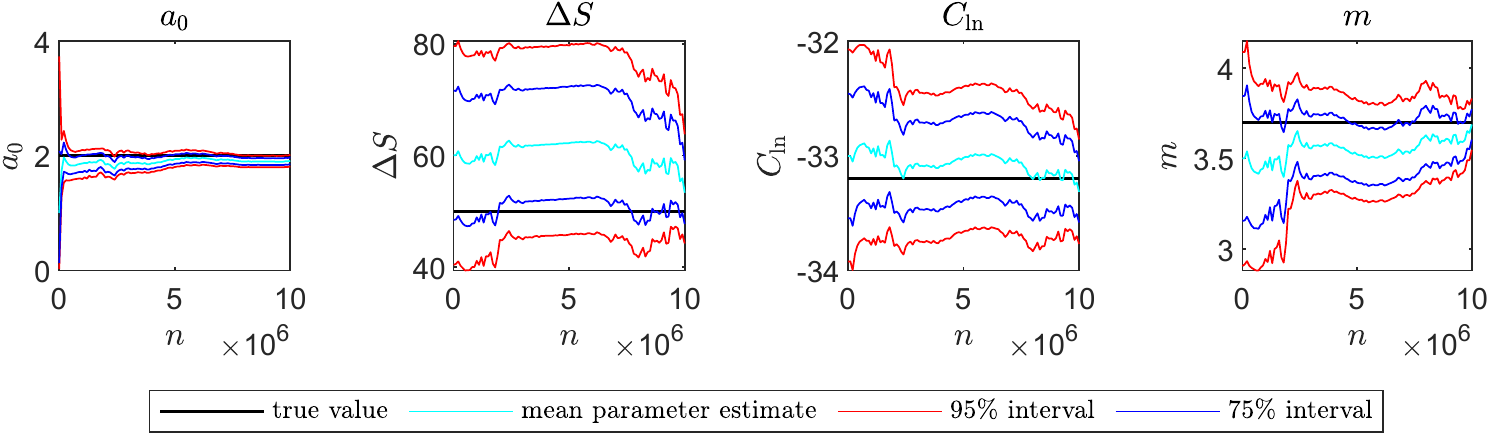}
	\caption{Reference posterior solution: mean and credible intervals for the sequence of posterior distributions $\pi_{\text{pos}}\left(\boldsymbol{\theta}|\mathbf{y}_{1:k}\right)$}
	\label{fig:cgm_ref_param_filt}
\end{figure}

\begin{figure}
	\centering
	\includegraphics[width=0.85\textwidth]{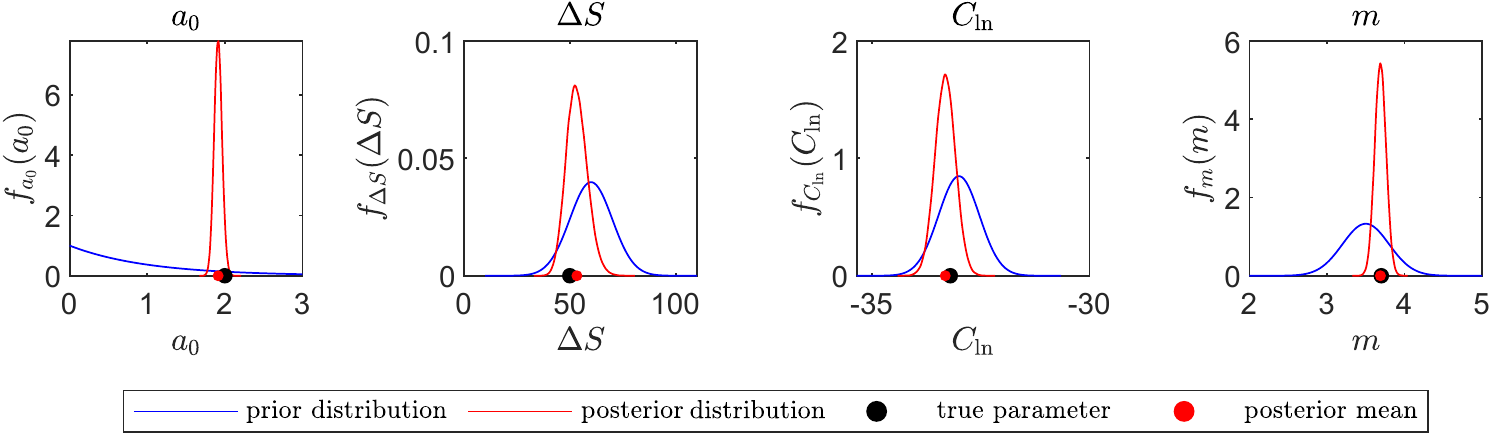}
	\caption{Reference final posterior: prior and single posterior distribution of interest $\pi_{\text{pos}}\left(\boldsymbol{\theta}|\mathbf{y}_{1:100}\right)$}
	\label{fig:cgm_ref_param_dist_final}
\end{figure}

Based on the generated measurements, the sequence of reference posterior distributions $\pi_{\text{pos}}\left(\boldsymbol{\theta}|\mathbf{y}_{1:k}\right)$ is obtained using the prior distribution as an envelope distribution for rejection sampling \citep{Smith_1992, Rubinstein_2016}. More specifically, for each of the 100 posterior distributions of interest $\pi_{\text{pos}}\left(\boldsymbol{\theta}|\mathbf{y}_{1:k}\right)$, $10^5$ independent samples are generated. The results of this reference posterior estimation of the four time-invariant model parameters are plotted in Figure \ref{fig:cgm_ref_param_filt}. With posterior samples, the reference filtered estimate of the crack length $a_n$ at each estimation step is also obtained via the model of Equation \eqref{eq:ODE_solution} and plotted in Figure \ref{fig:cgm_ref_state_filt}. In the left panel of this figure, the filtered state is plotted in logarithmic scale. In an off-line estimation, a single posterior density is of interest. One such reference posterior estimation result for the last estimation step, $\pi_{\text{pos}}\left(\boldsymbol{\theta}|\mathbf{y}_{1:100}\right)$, is plotted for illustration in Figure \ref{fig:cgm_ref_param_dist_final}.

\subsubsection{Comparative assessment of the investigated on-line and off-line filters}

\begin{figure}[!ht]
	\centering
	\includegraphics[width=0.85\textwidth]{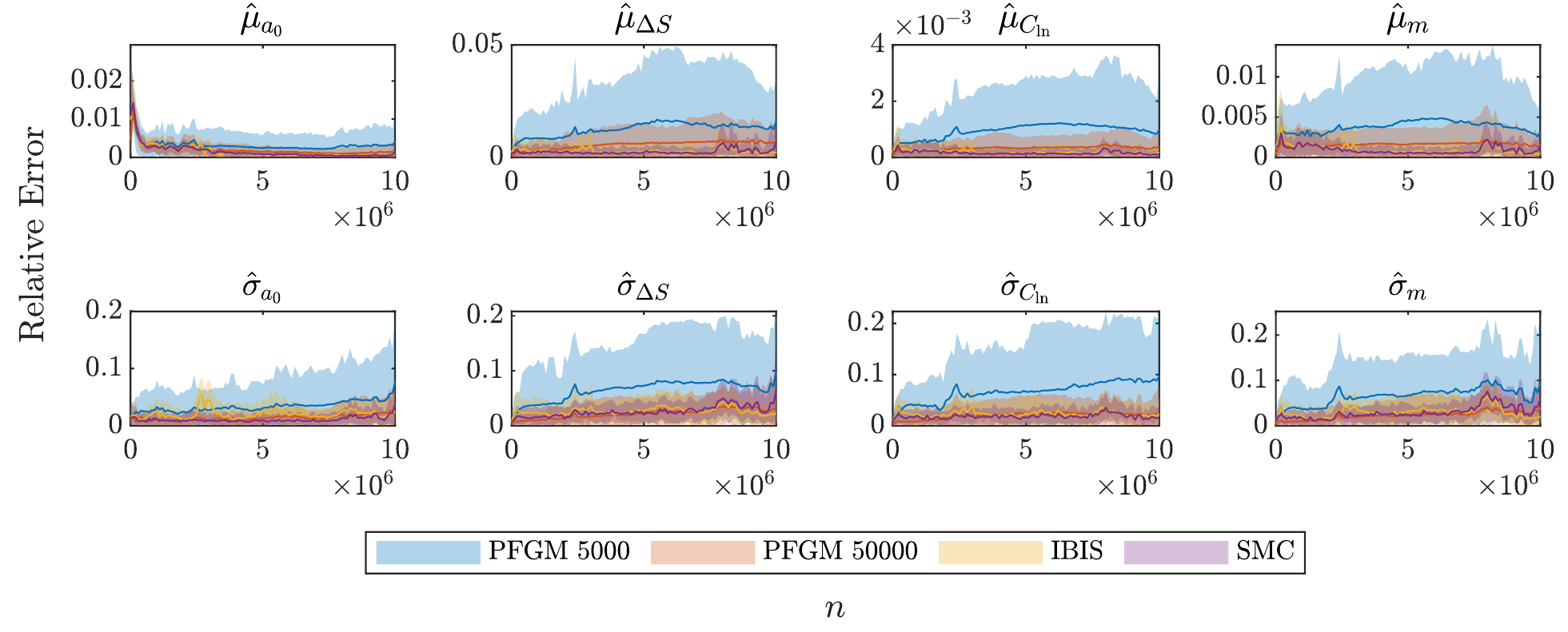}
	\vspace{-5 pt}
	\caption{Comparison of the relative error of the mean and standard deviation of the parameters evaluated for each filter. The solid lines show the mean and the shaded areas the 90\% credible intervals inferred from 50 repeated runs of each filter. In the horizontal axis, $n$ is the number of stress cycles}
	\label{fig:cgm_comp_filters_sep_err}
\end{figure}

We apply the PFGM filter with 5000 and 50000 particles, the IBIS filter with 5000 particles, and the SMC filter with 5000 particles for performing on-line and off-line time-invariant parameter estimation tasks. We evaluate the performance of each filter by taking the relative error of the estimated mean and standard deviation of each of the four parameters with respect to the reference posterior solution. For example, the relative error in the estimation of the mean of parameter $a_0$ at a certain estimation step $k$ is computed as $\abs{\frac{\mu_{a_0,k}-\hat{\mu}_{a_0,k}}{\mu_{a_0,k}}}$, where $\mu_{a_0,k}$ is the reference posterior mean from rejection-sampling (Section \ref{subsubsec:reference_post_CGM}), and $\hat{\mu}_{a_0,k}$ is the posterior mean estimated with each filter. Each filter is run 50 times, and the mean relative error of the mean and the standard deviation of each parameter, together with the 90\% credible intervals (CI), are obtained. These are plotted in Figure \ref{fig:cgm_comp_filters_sep_err}.

\begin{figure}
	\centering
	\includegraphics[width=0.85\textwidth]{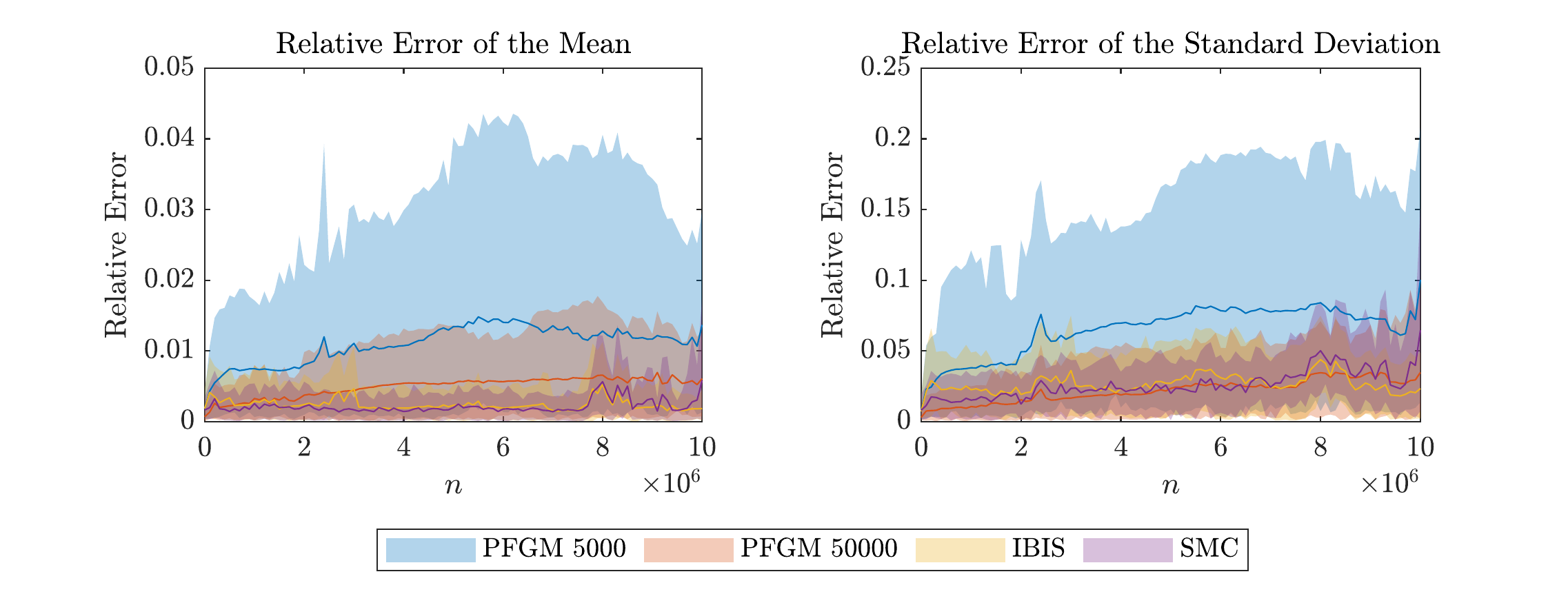}
	\vspace{-5pt}
	\caption{Comparison of the $L^2$ relative error norm of the mean and the standard deviation of the parameters evaluated for each filter. The solid lines show the mean and the shaded areas the 90\% credible intervals inferred from 50 repeated runs of each filter. In the horizontal axis, $n$ is the number of stress cycles}
	\label{fig:cgm_comp_filters_total_err}
\end{figure}

Figure \ref{fig:cgm_comp_filters_total_err} plots the $L^2$ relative error norm of the mean and the standard deviation of all four parameters, i.e., the quantity of equation \eqref{eq:L2_error} (here formulated for the mean at estimation step $k$)
\begin{equation}
	\sqrt{\frac{\sum_{i=1}^{d}\left(\mu_{i,k}-\hat{\mu}_{i,k}\right)^2}{\sum_{i=1}^{d}\left(\mu_{i,k}\right)^2}}
	\label{eq:L2_error}
\end{equation}
where $d$ is the dimensionality of the time-invariant parameter vector $\boldsymbol{\theta}$ (in this example $d=4$). More specifically, Figure \ref{fig:cgm_comp_filters_total_err} plots the mean and credible intervals of the $L^2$ relative error norm of the estimated mean and standard deviation, as obtained from 50 runs of each filter.

Figures \ref{fig:cgm_comp_filters_sep_err} and \ref{fig:cgm_comp_filters_total_err} reveal that, when all three filters are run with the same number of particles, the IBIS and SMC filters yield superior performance over PFGM. When the number of particles in the PFGM filter is increased to 50000, the PFGM filter performance is comparable to the one of the IBIS and SMC filters. In estimating the mean, the mean $L^2$ relative error norm obtained from the PFGM filter with 50000 particles is slightly larger than the corresponding error obtained from IBIS and SMC with 5000 particles, while the 90\% credible intervals of the PFGM filter estimation are still wider. In estimating the standard deviation, the PFGM filter with 50000 particles proves competitive.

Figures \ref{fig:cgm_comp_filters_sep_err} and \ref{fig:cgm_comp_filters_total_err} show the estimation accuracy of each filter when used for on-line inference, i.e., for estimating the whole sequence of 100 posterior distributions $\pi_{\text{pos}}\left(\boldsymbol{\theta}|\mathbf{y}_{1:k}\right)$, $k=1,\ldots,100$. The PFGM and IBIS filters, being intrinsically on-line filters, provide the whole posterior sequence with one run. On the other hand, the off-line SMC filter is run anew for each of the 100 required posterior estimations. Hence, Figures \ref{fig:cgm_comp_filters_sep_err} and \ref{fig:cgm_comp_filters_total_err} enclose the results of both the on-line and the off-line inference. If one is interested in the off-line estimation accuracy at a specific stress cycle $n$, one can simply consider a vertical ``cut" at $n$.

\begin{table}
	\caption{Average number of model evaluations for the fatigue crack growth model parameter estimation}
	\small
	\centering
	\begin{tabular}{|c|c|c|c|c|c|}\hline
		method &PFGM 5000&PFGM 50000&IBIS&SMC (final posterior)&SMC (all posteriors)\\\hline
		model evaluations&$5\times10^5$&$5\times10^6$&$3.4\times10^6$&$4.5\times10^6$&$1.9\times10^8$\\\hline
	\end{tabular}
	\label{table:Comp_cost_CGM} 
\end{table}

Table \ref{table:Comp_cost_CGM} documents the computational cost associated with each filter, expressed in the form of required model evaluations induced by calls of the likelihood function. By model we here refer to the model of Equation \eqref{eq:ODE_solution}, which is an analytical expression with negligible associated runtime. However, unlike the simple measurement equation that we have assumed in this example, in many realistic deterioration monitoring settings, the deterioration state cannot be measured directly (e.g., in vibration-based structural health monitoring \citep{Kamariotis_2022a}). In such cases, each deterioration model evaluation often entails evaluation of a finite element (FE) model, which has substantial runtime. It therefore appears appropriate to evaluate the filters' computational cost in terms of required model evaluations. The on-line PFGM filter with 5000 particles requires $5\times10^5$ model evaluations, and yields by far the smallest computational cost, while at the same time providing the solution to both on-line and off-line estimation tasks. However, it also yields the worst performance in terms of accuracy of the posterior estimates. Running the IBIS filter with 5000 particles, which performs MCMC move steps, leads to $3.4\times10^6$ model evaluations. Comparing this value against the $5\times10^5$ model evaluations required by the PFGM filter with 5000 particles for performing the same task distinctly shows the computational burden associated with MCMC move steps, which require a complete browsing of the whole measurement data set in estimating the acceptance probability. However, the IBIS filter also leads to enhanced estimation accuracy, which might prove significant when the subsequent tasks entail prognosis of the deterioration evolution, the structural reliability or the remaining useful lifetime, and eventually the predictive maintenance planning. Using 50000 particles, the PFGM filter performance increases significantly with a computational cost that is comparable to the IBIS filter with 5000 particles. For the off-line SMC algorithm, $4.5\times10^6$ model evaluations are required only for the task of estimating the final posterior density. The $1.9\times10^8$ model evaluations required by the SMC for obtaining the whole sequence of posteriors $\pi_{\text{pos}}\left(\boldsymbol{\theta}|\mathbf{y}_{1:k}\right)$, $k=1,\ldots,100$, clearly demonstrate that off-line MCMC techniques are unsuited to on-line estimation tasks.

\subsection{High-dimensional case study: Corrosion deterioration spatially distributed across beam}
\label{subsec:RF}

\begin{figure}[ht!]
	\centering
	\includegraphics[width=0.75\textwidth]{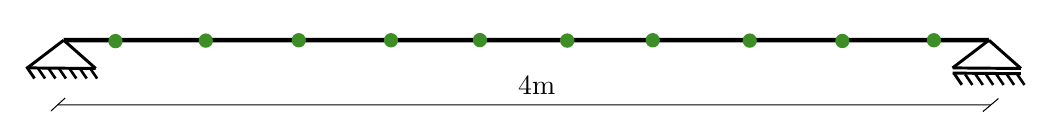}
	\vspace{-5pt}
	\caption{Structural beam subjected to spatially and temporally varying corrosion deterioration. The deterioration process is monitored from sensors deployed at specific sensor locations (in green)}
	\label{fig:beam_corrosion}
\end{figure}

As a second case study, we employ the deterioration model of Equation \eqref{eq:RF_deterioration}, which describes the spatially and temporally varying corrosion deterioration across the structural beam shown in Figure \ref{fig:beam_corrosion}.
\begin{equation}
	D(t)=At^B, \qquad t=0, \dots,50
	\label{eq:RF_deterioration}
\end{equation}
$A$ is a random field modeling the deterioration rate, while $B$ is a random field related to the nonlinearity effect of the deterioration process in terms of a power law in time. The corrosion deterioration $D(t)$ is therefore also a spatial random field. 

A random field, by definition, contains an infinite number of random variables, and must therefore be discretized \citep{Vanmarcke_2010}. One of the most common methods for discretization of random fields is the midpoint method \citep{DERKIUREGHIAN_1988}, whereby the domain is discretized in $m$ elements, and the two random fields can be approximated by using the random variables that correspond to the values of the random fields at the discrete points in the domain (the midpoints of each element). In that case, the uncertain time-invariant deterioration model parameter vector is $\boldsymbol{\theta} = \left[A_1,\dots, A_m, B_1, \dots, B_m\right]$, where $A_i, B_i, i = 1, \ldots, m$ are the random variables corresponding to the midpoint of the $i$-th element.

We assume that noisy measurements of the corrosion deterioration state $D_{t,l}$ at time $t$ and at certain locations $l$ of the beam are obtained sequentially (summarized in one measurement per year) from  $n_{l}$ sensors deployed at these locations ($n_{l} = 10$ sensor locations are shown in Figure \ref{fig:beam_corrosion}). The measurement Equation \eqref{eq:RF_measurement}, describing the corrosion measurement at time $t$ and sensor location $l$, assumes a multiplicative measurement error, $\exp \left(\epsilon_{t,l}\right)$.
\begin{equation}
	y_{t,l} = D_{t,l}\left(\boldsymbol{\theta}\right) \exp \left(\epsilon_{t,l}\right) = A_{i_l} t^{B_{i_l}} \exp \left(\epsilon_{t,l}\right), 
	\label{eq:RF_measurement}
\end{equation}
where ${i_l}$ returns the discrete element number of the midpoint discretization within which the measurement location l lies.  
Table \ref{tab:par_RF} shows the prior distribution model for the two random fields of the deterioration model of Equation \eqref{eq:RF_deterioration} and the assumed probabilistic model of the multiplicative measurement error. Since $A$ models a lognormal random field, $\ln(A)$ follows the normal distribution. For both random fields $\ln(A)$ and $B$, the exponential correlation model with correlation length of 2m is applied \citep{Sudret_2000}.

\begin{table}[ht!]
	\caption{Prior distribution model for the corrosion deterioration model parameters and the measurement error}
	\small
	\begin{tabular*}{\textwidth}{@{\extracolsep{\fill}} c c c c c}
		\hline
		\textbf{Parameter} & \textbf{Distribution} & \textbf{Mean} & \textbf{Standard Deviation} & \textbf{Corr. length (m) }\\
		\hline
		$A$ & Lognormal & $0.8$ & $0.24$ & $2$\\
		$B$ & Normal & $0.8$ & $0.12$ & $2$\\
		$\exp \left( \epsilon_{t,l} \right)$ & Lognormal & $1.0$ & $0.101$& -\\
		\hline
	\end{tabular*}
	\label{tab:par_RF}
\end{table}
The goal is to update the time-invariant deterioration model parameters $\boldsymbol{\theta} = \left[A_{1},\dots, A_{m}, B_1, \dots, B_m\right]$ given sequential noisy corrosion measurements $y_{t,l}$ from $n_l$ deployed sensors. The dimensionality of the problem is $d = 2\times m$. Hence, the more elements in the midpoint discretization, the higher the dimensionality of the parameter vector.

The main goal of this second case study is to investigate the effect of the problem dimensionality and the amount of sensor information on the posterior results obtained with each filter. We choose the following three midpoint discretization schemes:
\begin{enumerate}[noitemsep,topsep=0pt]
	\item $m=25$ elements: $d=50$ time-invariant parameters to estimate.
	\item $m=50$ elements: $d=100$ time-invariant parameters to estimate.
	\item $m=100$ elements: $d=200$ time-invariant parameters to estimate.
\end{enumerate}
Furthermore, we choose the following three potential sensor arrangements: 
\begin{enumerate}[noitemsep,topsep=0pt]
	\item $n_l=2$ sensors (the $4^{th}$ and $7^{th}$ sensors of Figure \ref{fig:beam_corrosion}).
	\item $n_l=4$ sensors (the $1^{st}, 4^{th}, 7^{th}$ and $10^{th}$ sensors of Figure \ref{fig:beam_corrosion}).
	\item $n_l=10$ sensors of Figure \ref{fig:beam_corrosion}.
\end{enumerate}
We therefore study nine different cases of varying problem dimensionality and number of sensors.

\subsubsection{Markovian state-space representation for application of on-line filters}
\label{subsubsec:state_space_RF}
A Markovian state-space representation of the deterioration process is required for application of on-line filters. The dynamic and measurement equations are shown in Equation \eqref{eq:state_space_RF}. The measurement equation is written in the logarithmic scale. Time $t$ is discretized in yearly estimation time steps $k$, i.e., $k=1,\dots,50$, and the subscript $l=1,\dots,n_l$ corresponds to the sensor location.
\begin{equation}
	\begin{split}
		\boldsymbol{\theta}_k &= \boldsymbol{\theta}_{k-1} \\
		\ln \left(y_{k,l}\right) &= \ln \left(D_{k,l}(\boldsymbol{\theta}_k)\right) + \epsilon_{k,l} \Rightarrow 
		\ln \left(y_{k,l}\right) = \ln \left(A_{k,i_l}\right) + B_{k,i_l} \ln \left(t_k\right) +\epsilon_{k,l}
		\label{eq:state_space_RF}
	\end{split}
\end{equation}
In the logarithmic scale, both the dynamic and measurement equations are linear functions of Gaussian random variables. For reasons explained in Section \ref{sec:algorithms}, the subscript $k$ in $\boldsymbol{\theta}_k$ is dropped in the following.

\subsubsection{Underlying ``true" realization}
\label{subsubsec:under_true_RF}

\begin{figure}[ht!]
	\centering
	\begin{subfigure}{0.48\textwidth}
		\centering
		\includegraphics[width=0.95\linewidth]{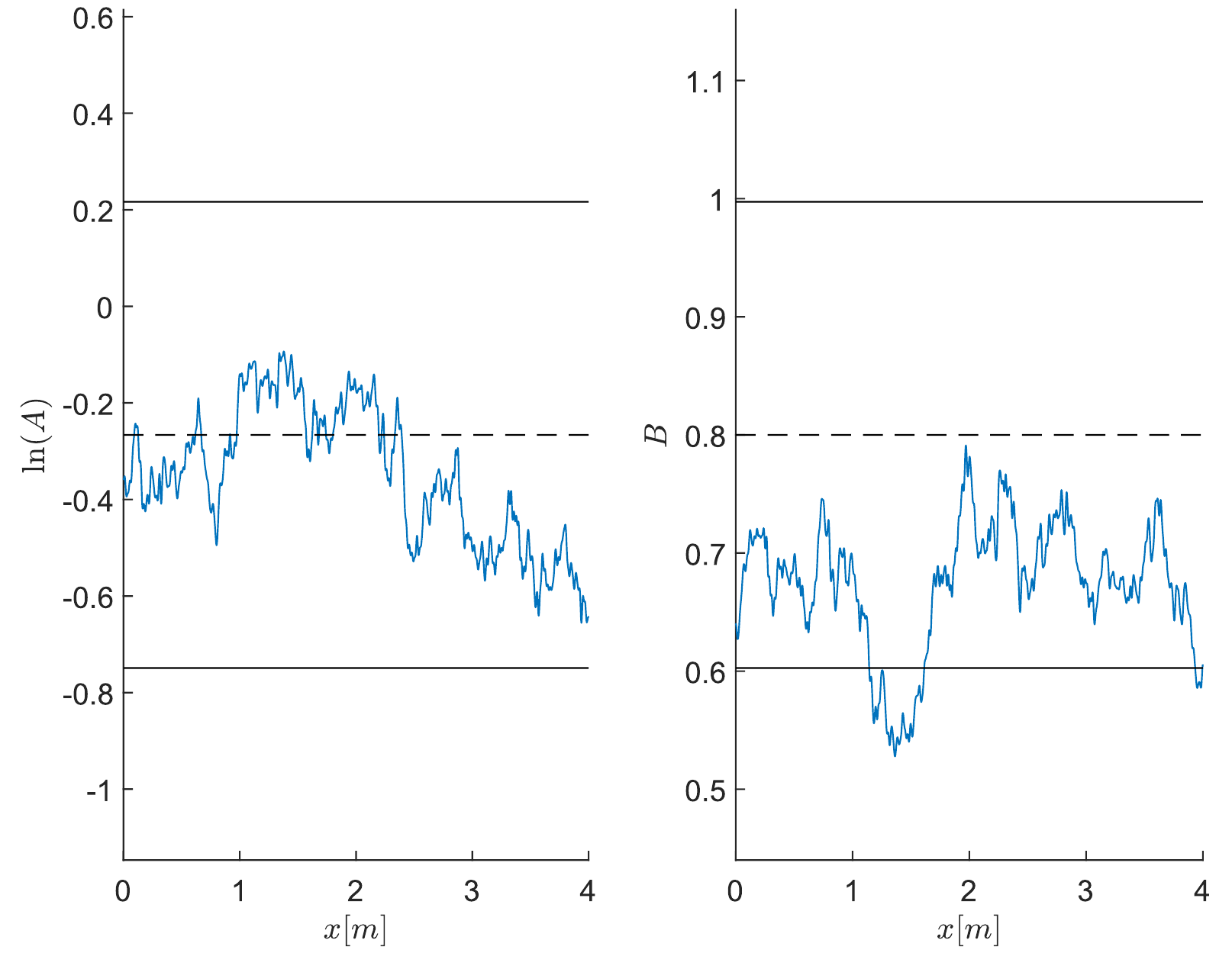}
	\end{subfigure}%
	~ 
	\begin{subfigure}{0.48\textwidth}
		\centering
		\includegraphics[width=0.95\linewidth]{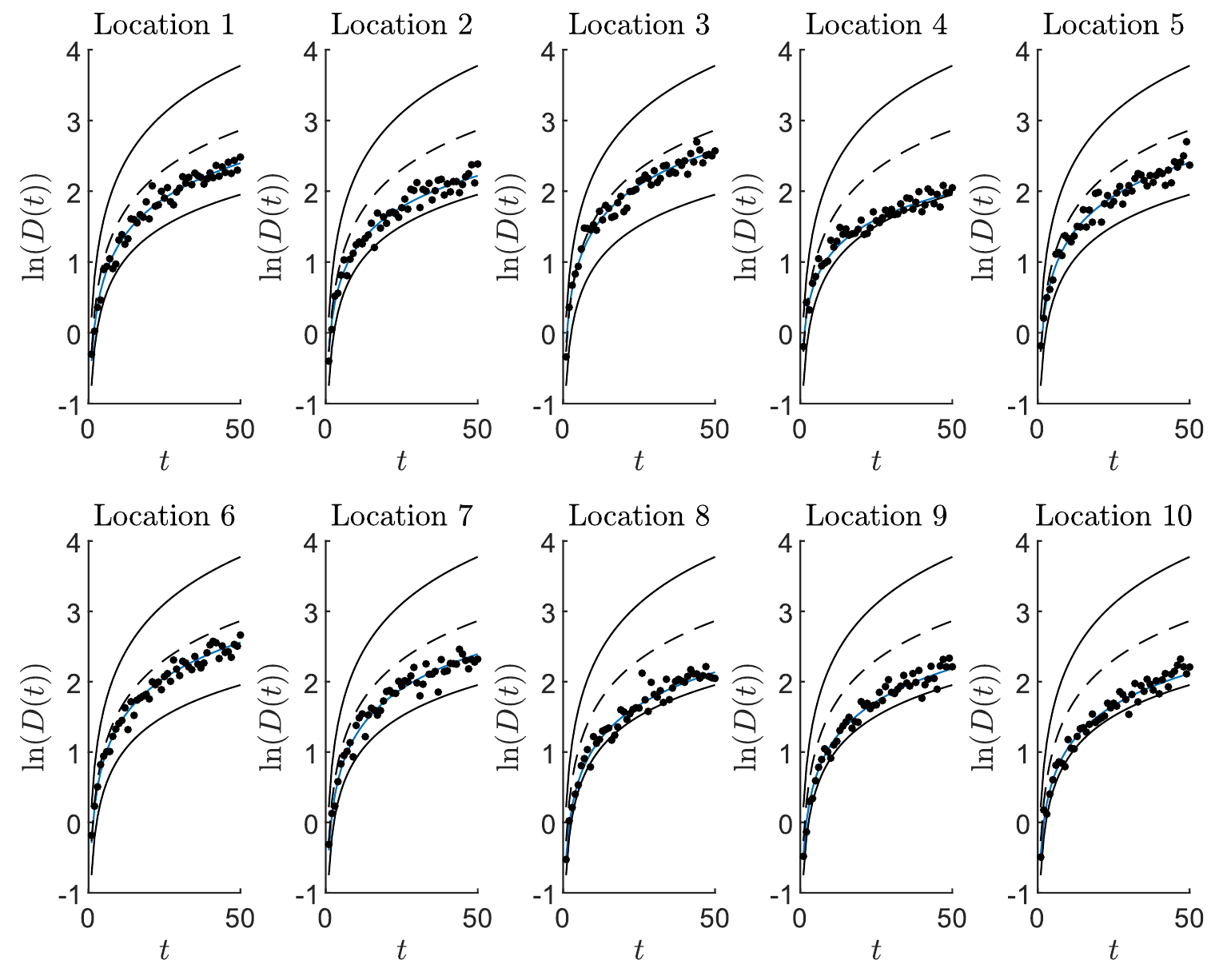}
	\end{subfigure}
	\caption{Left: the blue solid line plots the underlying ``true" realization of $\ln(A)$ and $B$ created using the KL expansion. Right: the blue solid line plots the underlying ``true" realization of $\ln\left(D(t)\right)$ at 10 specific sensor locations and the corresponding synthetic sensor monitoring data are scattered in black. In both figures, the black dashed lines plot the prior mean and the black solid lines the prior 90\% credible intervals}
	\label{fig:KL}
\end{figure}

To generate a high-resolution underlying ``true" realization of the two random fields $A$ and $B$, and the corresponding synthetic monitoring data set, we employ the Karhunen-Loeve (KL) expansion \citep{Sudret_2000} using the first 400 KL modes. These realizations are shown in the left panel of Figure \ref{fig:KL}. Given these $A$ and $B$ realizations, the underlying ``true" realizations of the deterioration process at ten specific beam locations are generated, which correspond to the ten potential sensor placement locations shown in Figure \ref{fig:beam_corrosion}. Subsequently, a synthetic corrosion sensor measurement data set (one measurement per year) at these 10 locations is generated from the measurement Equation \eqref{eq:RF_measurement}. These are shown in the right panel of figure \ref{fig:KL}. The KL expansion is used for the sole purpose of generating the underlying ``truth". 

\subsubsection{Reference posterior solution}
\label{subsubsec:ref_post_RF}

\begin{figure}[ht!]
	\centering
	\begin{subfigure}{0.48\textwidth}
		\centering
		\includegraphics[width=0.95\linewidth]{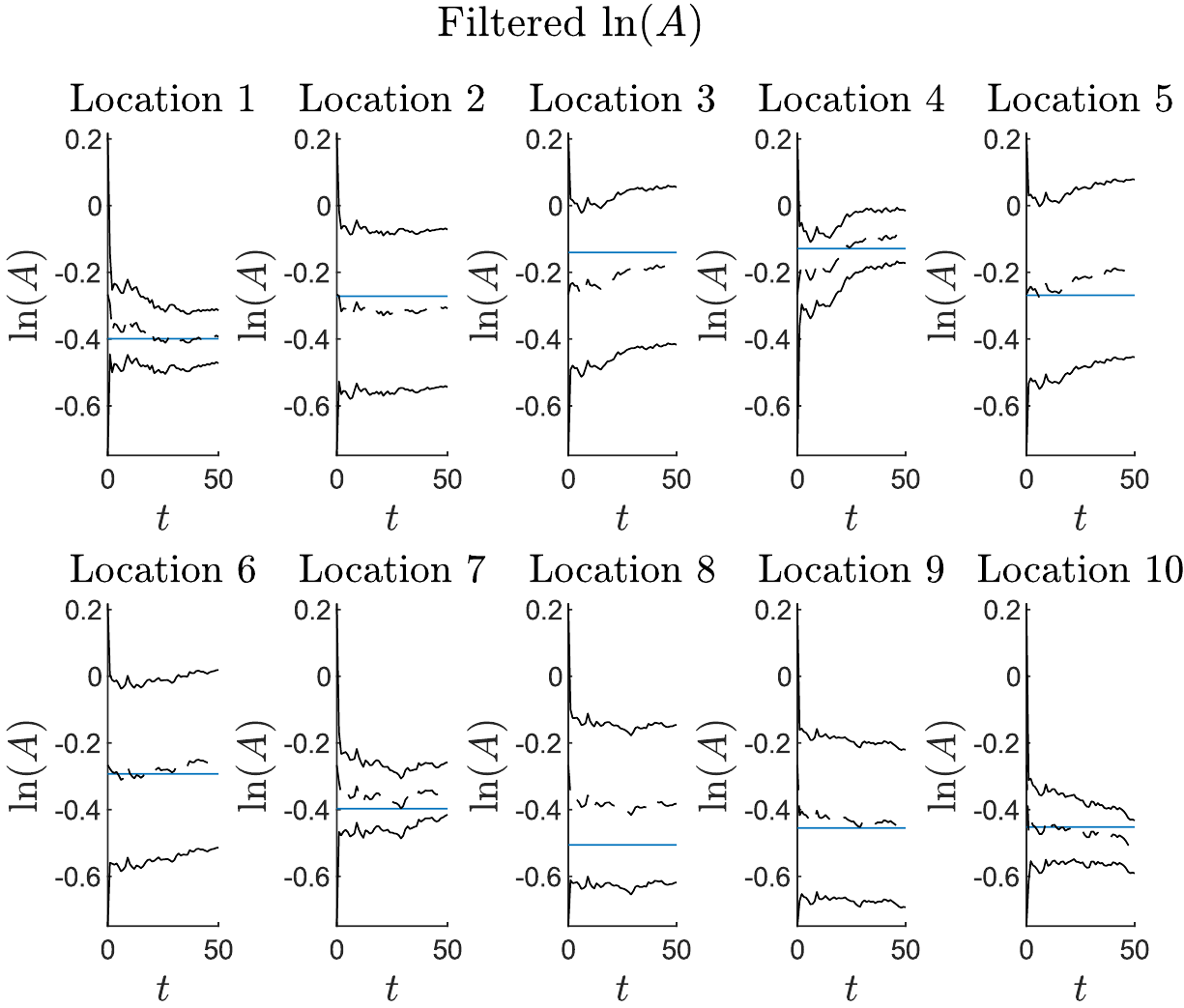}
	\end{subfigure}%
	~ 
	\begin{subfigure}{0.48\textwidth}
		\centering
		\includegraphics[width=0.95\linewidth]{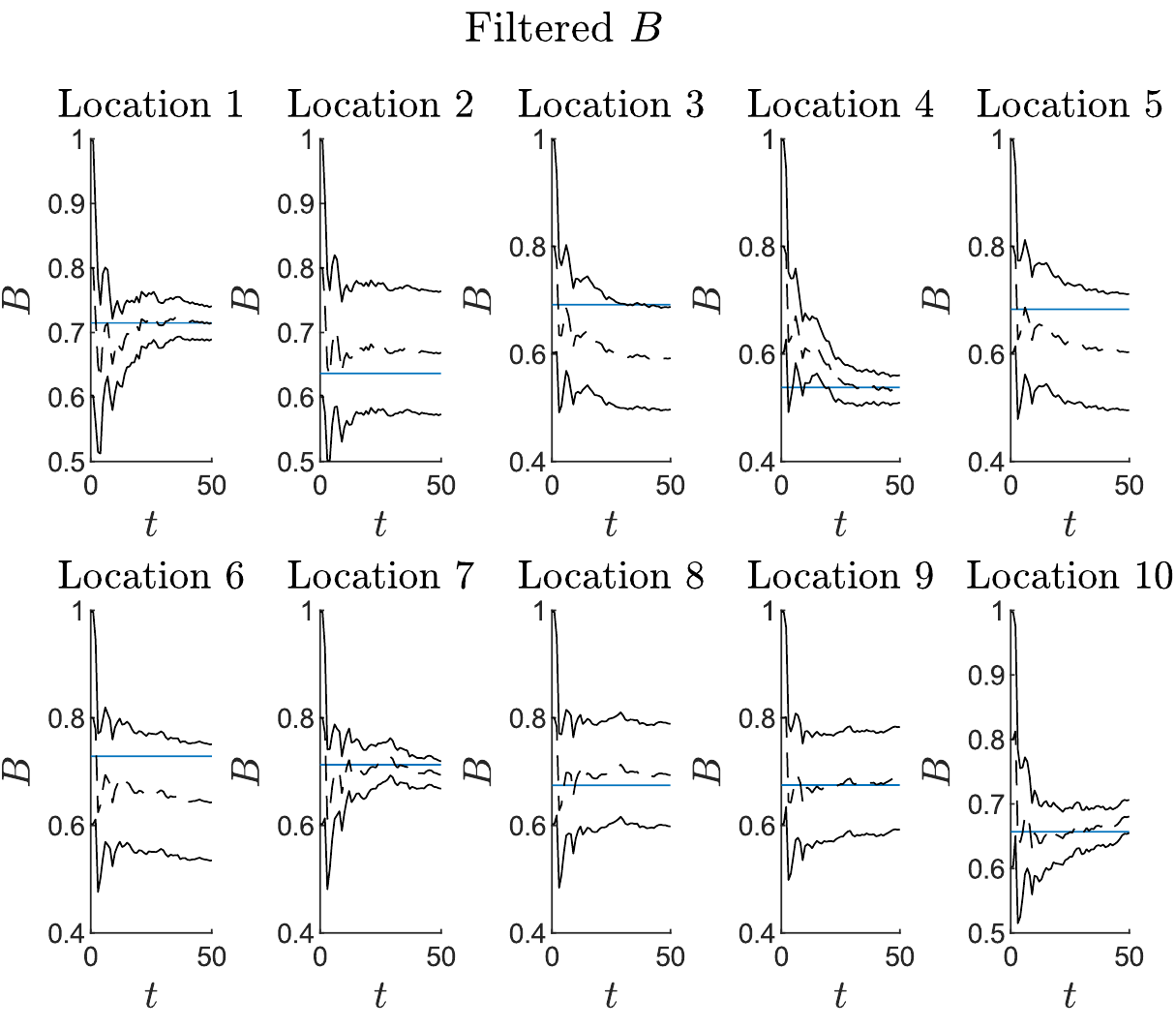}
	\end{subfigure}
	\caption{Case with $m=25, n_l=4$: reference on-line posterior solution at 10 locations across the beam obtained with the Kalman filter. The solid blue horizontal line represents the underlying ``true" values of $ln(A)$ and $B$ at these locations. The black dashed lines plot the posterior mean and the black solid lines the posterior 90\% credible intervals. Locations 1,4,7,10 correspond to the four assumed sensor placement locations}
	\label{fig:KF_ref}
\end{figure}

For the investigated linear Gaussian state space representation of Equation \ref{eq:state_space_RF}, we create reference on-line posterior solutions for each of the nine considered cases by applying the Kalman filter (KF) \citep{Kalman_1960}, which is the closed form solution to the Bayesian filtering equations. The process noise covariance matrix in the KF equations is set equal to zero. The linear Gaussian nature of the chosen problem ensures existence of an analytical reference posterior solution obtained with the KF. One such reference on-line posterior solution for the case described by $m=25$ elements ($d=50$) and $n_l=4$ sensors is shown in Figure \ref{fig:KF_ref}. 

\subsubsection{Comparative assessment of the investigated on-line and off-line filters}

\begin{figure}[!ht]
	\centering
	\includegraphics[width=0.75\textwidth]{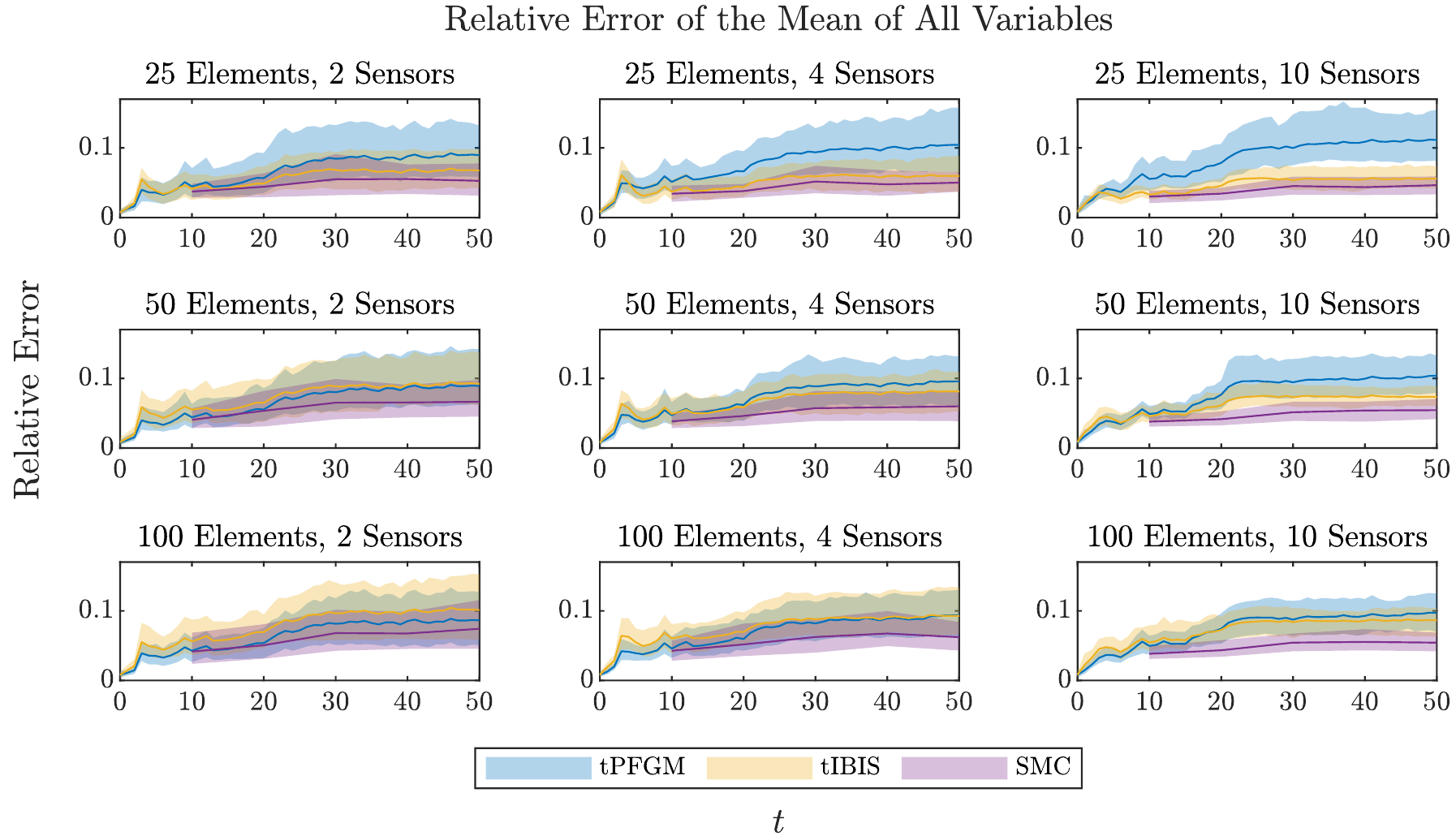}
	\caption{Comparison of the $L^2$ relative error norm of the means of the parameters evaluated for each filter. The solid lines show the mean and the shaded areas the 90\% credible intervals inferred from 50 repeated runs of each filter}
	\label{fig:RF_comp_filters_mean}
\end{figure}

\begin{figure}[!ht]
	\centering
	\includegraphics[width=0.75\textwidth]{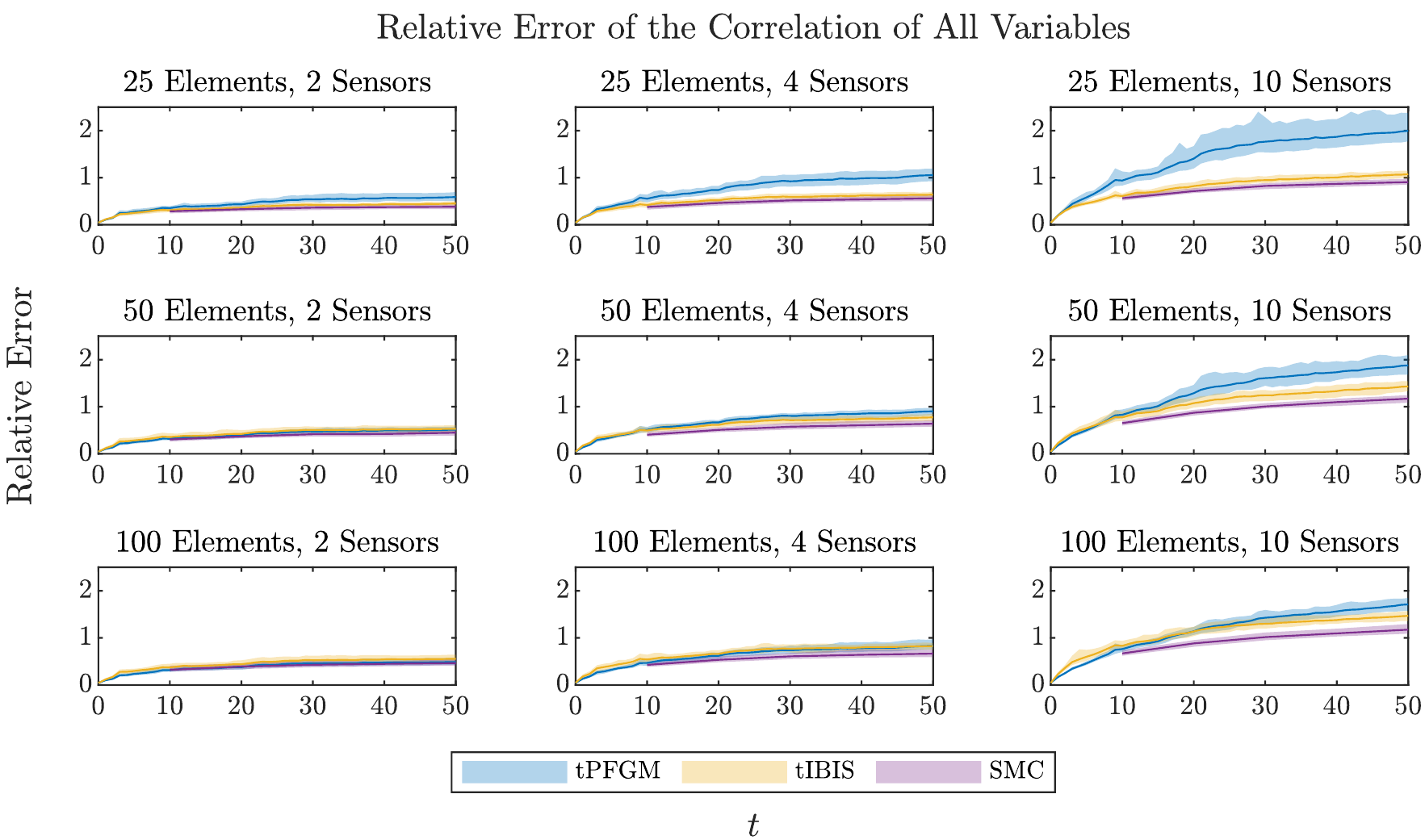}
	\caption{Comparison of the $L^2$ relative error norm of the correlation coefficients of the parameters evaluated for each filter. The solid lines show the mean and the shaded areas the 90\% credible intervals inferred from 50 repeated runs of each filter}
	\label{fig:RF_comp_filters_corr}
\end{figure}

\begin{figure}[!ht]
	\centering
	\includegraphics[width=0.75\textwidth]{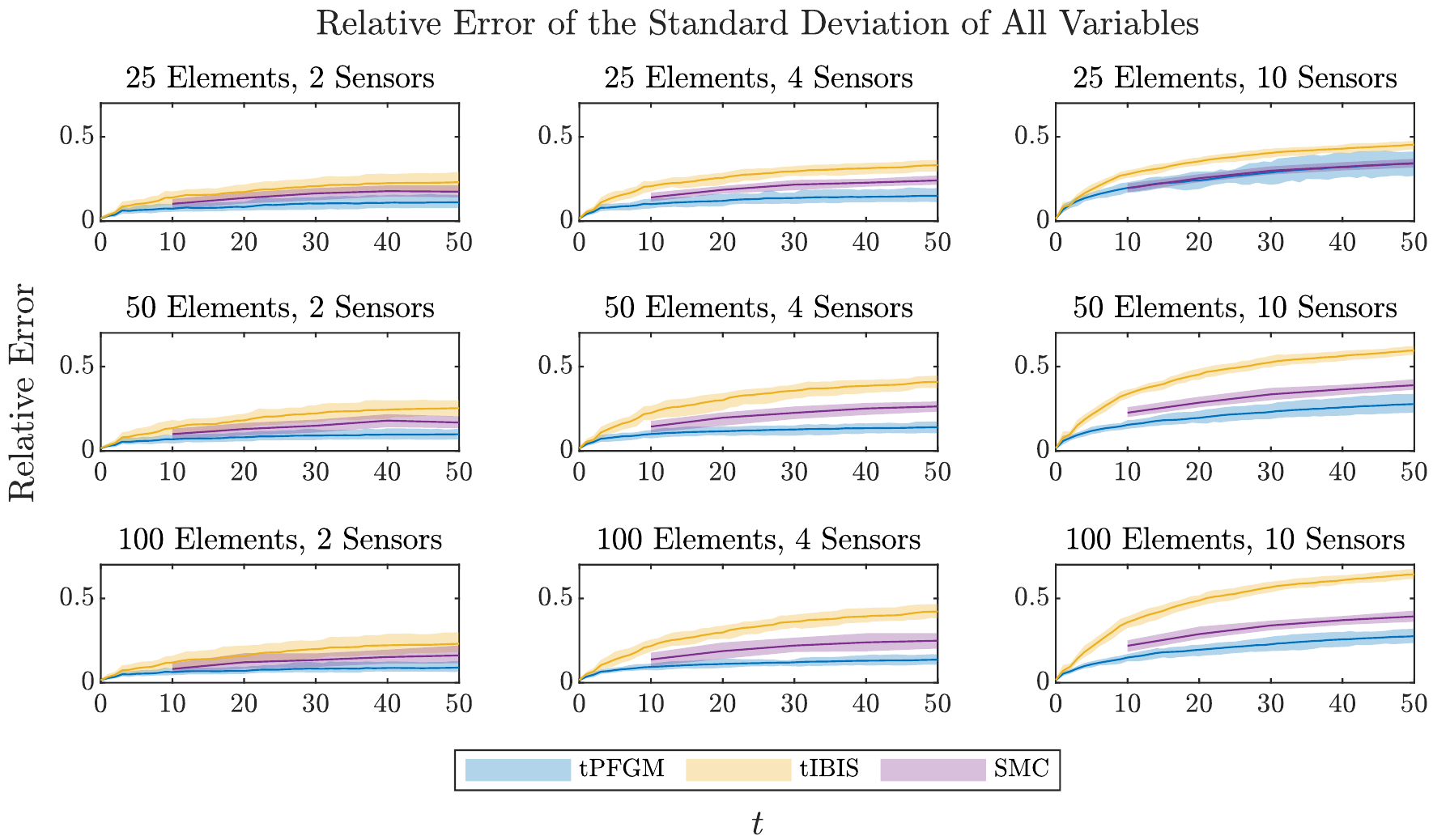}
	\caption{Comparison of the $L^2$ relative error norm of the standard deviations of the parameters evaluated for each filter. The solid lines show the mean and the shaded areas the 90\% credible intervals inferred from 50 repeated runs of each filter}
	\label{fig:RF_comp_filters_std}
\end{figure}

We apply the tPFGM filter, the tIBIS filter, and the SMC filter, all with $N_{\text{par}}$=2000 particles, for estimating the time-invariant parameter vector $\boldsymbol{\theta}$.  For each of the nine cases of varying problem dimensionality and number of sensors described above, we compute the $L^2$ relative error norm of the estimated means, correlation coefficients, and standard deviations of the parameters with respect to the corresponding KF reference posterior solution, i.e., we estimate a quantity as in Equation \eqref{eq:L2_error} for all estimation steps $k=1,\dots,50$. In Figures \ref{fig:RF_comp_filters_mean}, \ref{fig:RF_comp_filters_corr},  \ref{fig:RF_comp_filters_std} we plot the mean and credible intervals of these relative errors as obtained from 50 different runs. The off-line SMC filter, which does not provide the on-line solution within a single run, is run anew for estimating the single posterior density of interest at years 10, 20, 30, 40, 50, and in between, the relative error is linearly interpolated. Although each of the nine panels in the figures corresponds to a different case with a different underlying KF reference solution, their y axes have the same scaling. Table \ref{tab:hd_model_eval} documents the computational cost of each filter in each considered case, measured by average number of evaluations of the model of Equation \eqref{eq:RF_deterioration}.

Figures \ref{fig:RF_comp_filters_mean} and \ref{fig:RF_comp_filters_corr} show that the off-line IMH-GM-based SMC filter yields the best performance in estimating the KF reference posterior mean and correlation, for all nine considered cases, while at the same time producing the narrowest credible intervals. Comparison of the relative errors obtained with the SMC and tIBIS filters reveals that, although they are both reliant on the IMH-GM MCMC move step, the on-line tIBIS filter leads to larger estimation errors. The on-line tPFGM and tIBIS filters generate quite similar results in estimating the reference posterior mean and correlation, thus rendering the benefit of the MCMC move step in tIBIS unclear, except in cases with more sensors and lower parameter dimension. Figures \ref{fig:RF_comp_filters_mean} and \ref{fig:RF_comp_filters_corr} reveal a slight trend, indicating that for fixed dimensionality, availability of more sensors (i.e., stronger information content in the likelihood function) leads to a slight decrease in the relative errors when using the SMC and tIBIS filters, whereas the opposite trend can be identified for the tPFGM filter. Increasing problem dimensionality (for fixed number of sensors) does not appear to have strong influence on the posterior results in any of the columns of Figures \ref{fig:RF_comp_filters_mean}, \ref{fig:RF_comp_filters_corr} and \ref{fig:RF_comp_filters_std}, a result that initially appears puzzling.


\begin{table}[ht!]
	\fontsize{7pt}{12pt}\selectfont
	\caption{Average number of model evaluations for the high-dimensional case study. For the SMC, the required model evaluations for obtaining the single final posterior density are reported.}
	\begin{tabular*}{\textwidth}{@{\extracolsep{\fill}} |c|c|c|c|c|c|c|c|c|c| }
		\hline
		elements  & \multicolumn{3}{c|}{$25$} & \multicolumn{3}{c|}{$50$} & \multicolumn{3}{c|}{$100$}\\
		\hline
		sensors & $2$ & $4$ & $10$ & $2$ & $4$ & $10$ & $2$ & $4$ & $10$\\
		\hline
		tPFGM & 129,480 & 154,000 & 194,440 & 129,440 & 155,560 & 195,760 & 130,480 & 157,120 & 199,040\\
		\hline
		tIBIS & 602,400 & 1,038,440 & 1,878,000 & 603,240 & 1,049,400 & 1,909,880 & 567,720 & 1,017,280 & 1,876,240\\
		\hline
		SMC & 1,130,000 & 1,596,000 & 2,298,000 & 1,108,000 & 1,582,000 & 2,250,000 & 1,100,000 & 1,504,000 & 2,150,000\\
		\hline
	\end{tabular*}
	\label{tab:hd_model_eval}
\end{table}

Figure \ref{fig:RF_comp_filters_std} conveys that the tPFGM filter, which entirely depends on the GMM posterior approximation, induces the smallest relative errors for the estimation of the standard deviation of the parameters in all considered cases. This result reveals a potential inadequacy of the single application of the IHM-GM kernel for the move step within the tIBIS and SMC filters in properly exploring the space of $\boldsymbol{\theta}$. In all 50 runs of the tIBIS and SMC filters, the standard deviation of the parameters is consistently underestimated compared to the reference, unlike when applying the tPFGM filter. 

\begin{figure}[!htp]
	\centering
	\includegraphics[width=0.72\textwidth]{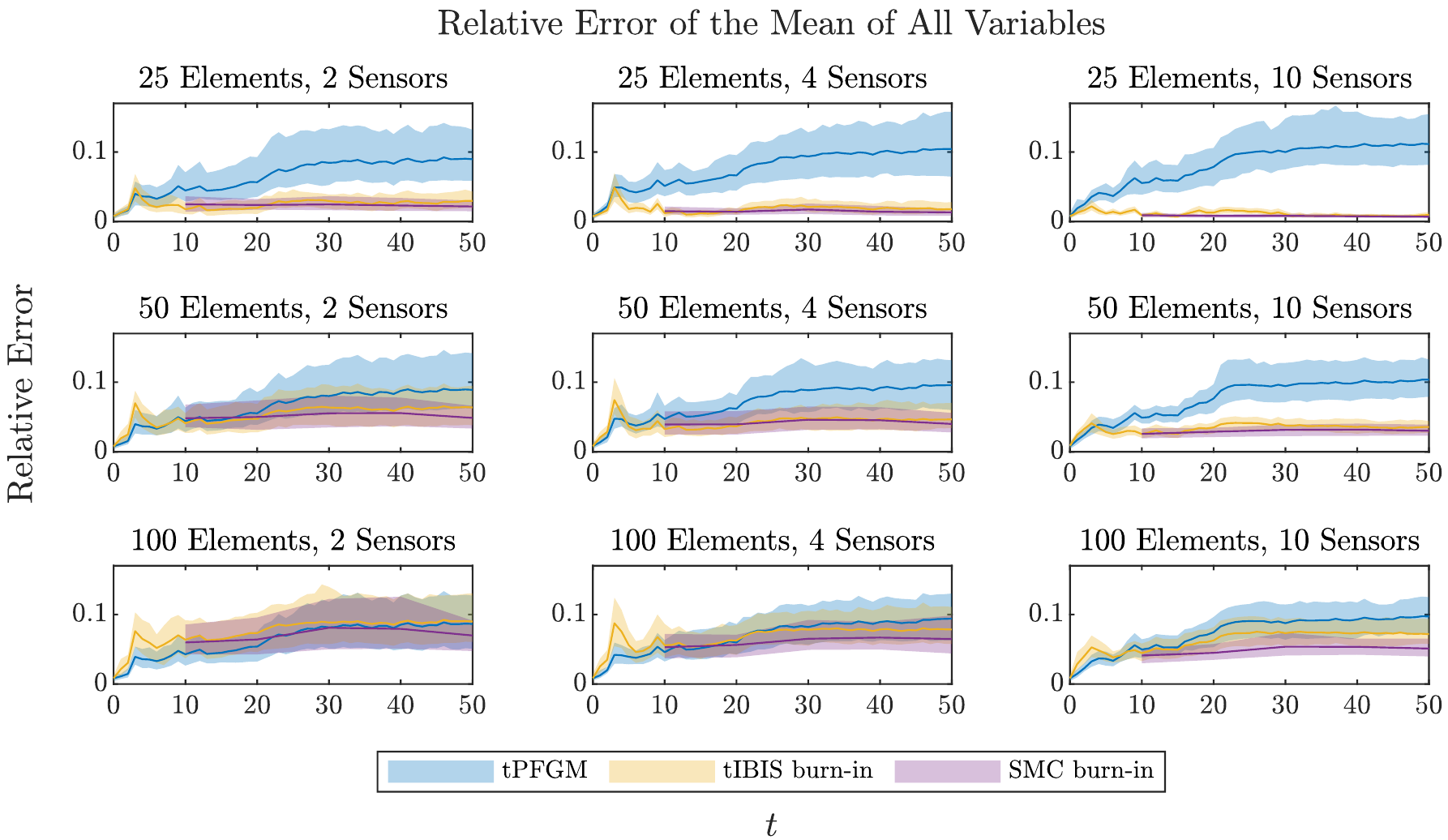}
	\vspace{-0.2cm}
	\caption{Comparison of the $L^2$ relative error norm of the mean of the parameters evaluated for each filter. The solid lines show the mean and the shaded areas the 90\% credible intervals inferred from 50 repeated runs of each filter. Burn-in $n_\text{B}$=5}
	\label{fig:RF_comp_filters_mean_burn_in}
\end{figure}

\begin{figure}[!htp]
	\centering
	\includegraphics[width=0.72\textwidth]{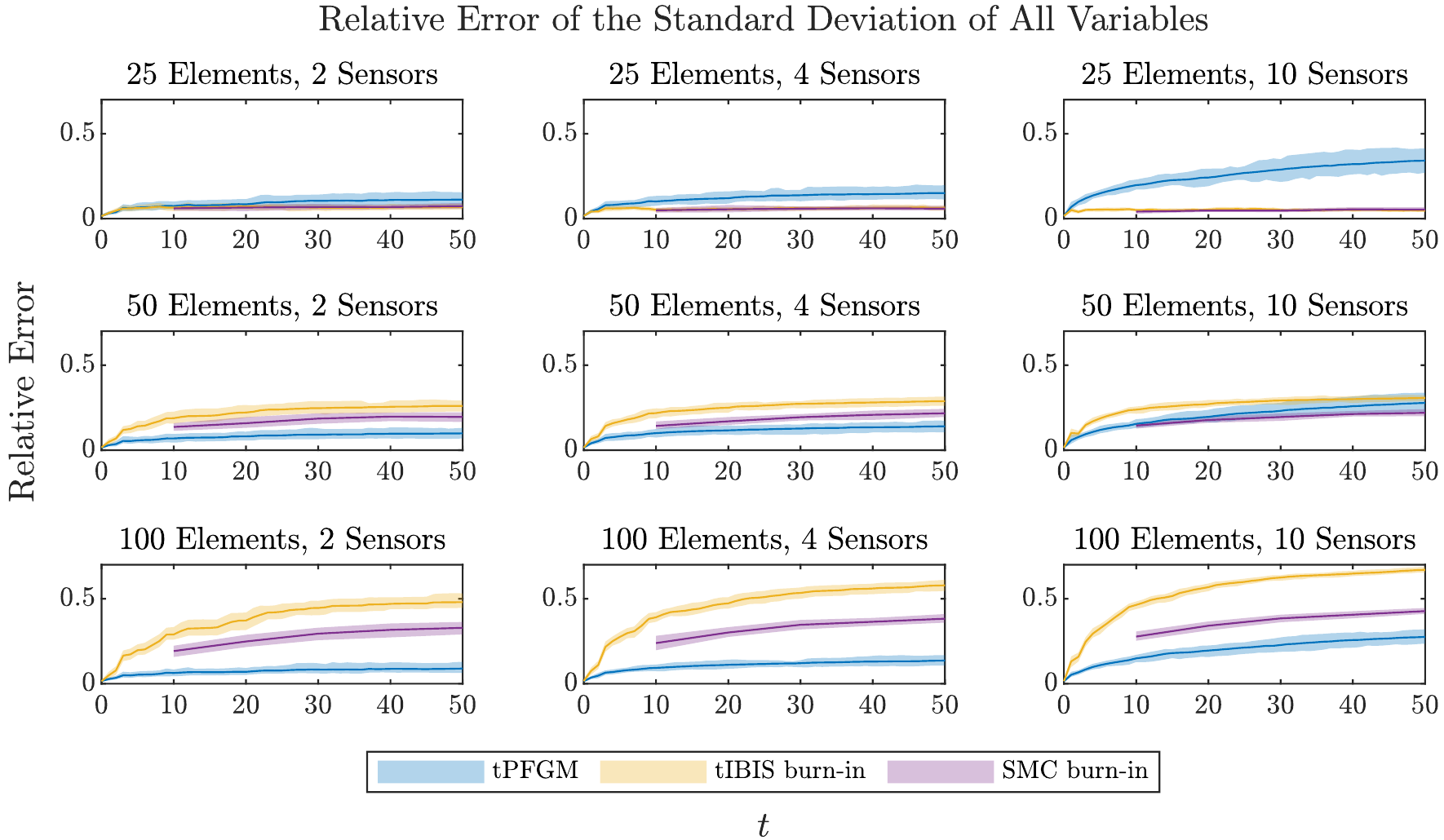}
	\vspace{-0.2cm}
	\caption{Comparison of the $L^2$ relative error norm of the standard deviation of the parameters evaluated for each filter. The solid lines show the mean and the shaded areas the 90\% credible intervals inferred from 50 repeated runs of each filter. Burn-in $n_\text{B}$=5}
	\label{fig:RF_comp_filters_std_burn_in}
\end{figure}

\begin{figure}[h]
	\centering
	\begin{subfigure}{0.35\textwidth}
		\centering
		\includegraphics[width=0.9\linewidth]{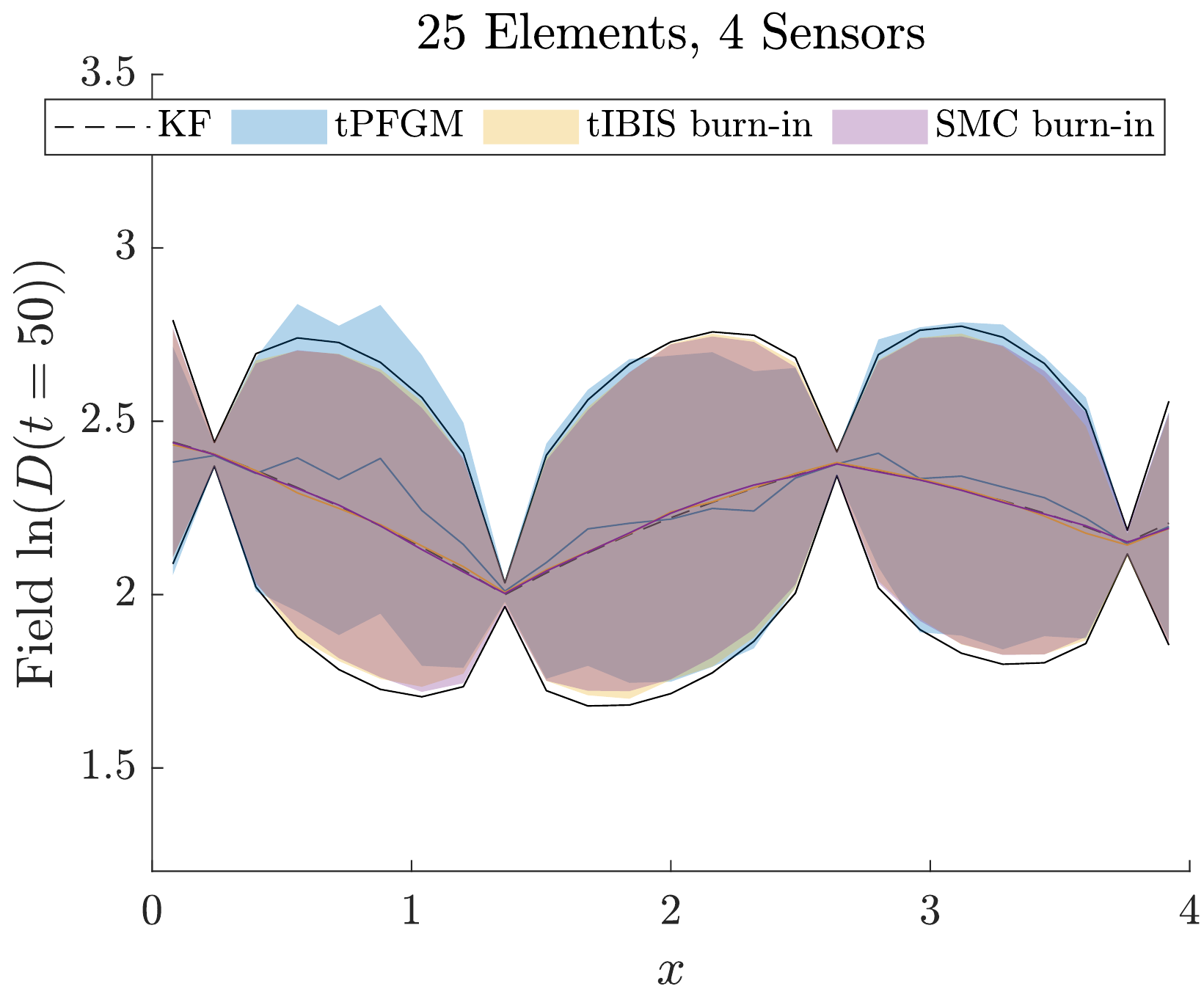}
	\end{subfigure}%
	~ 
	\begin{subfigure}{0.35\textwidth}
		\centering
		\includegraphics[width=0.9\linewidth]{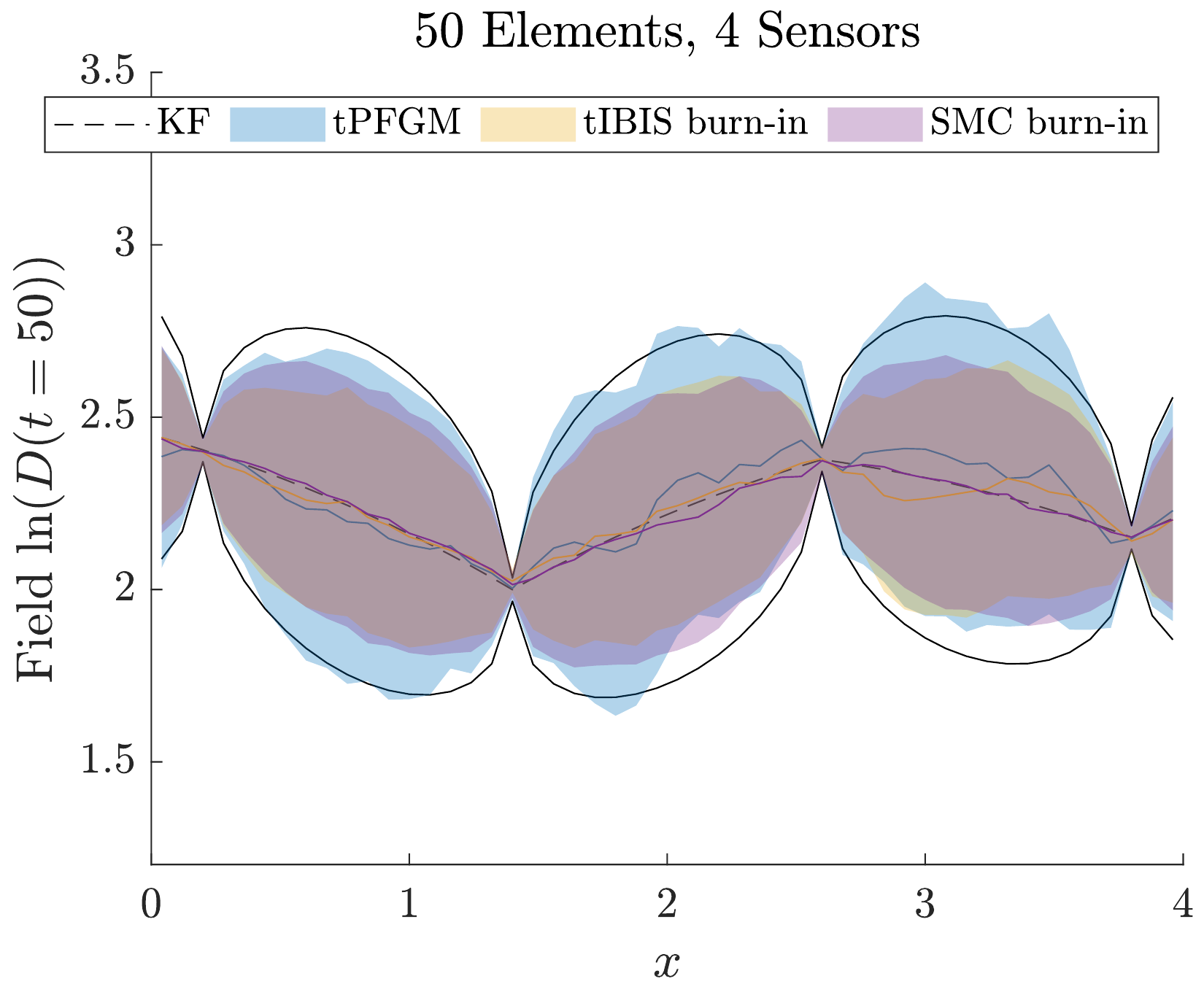}
	\end{subfigure}
	~ 
	\begin{subfigure}{0.35\textwidth}
		\centering
		\includegraphics[width=0.9\linewidth]{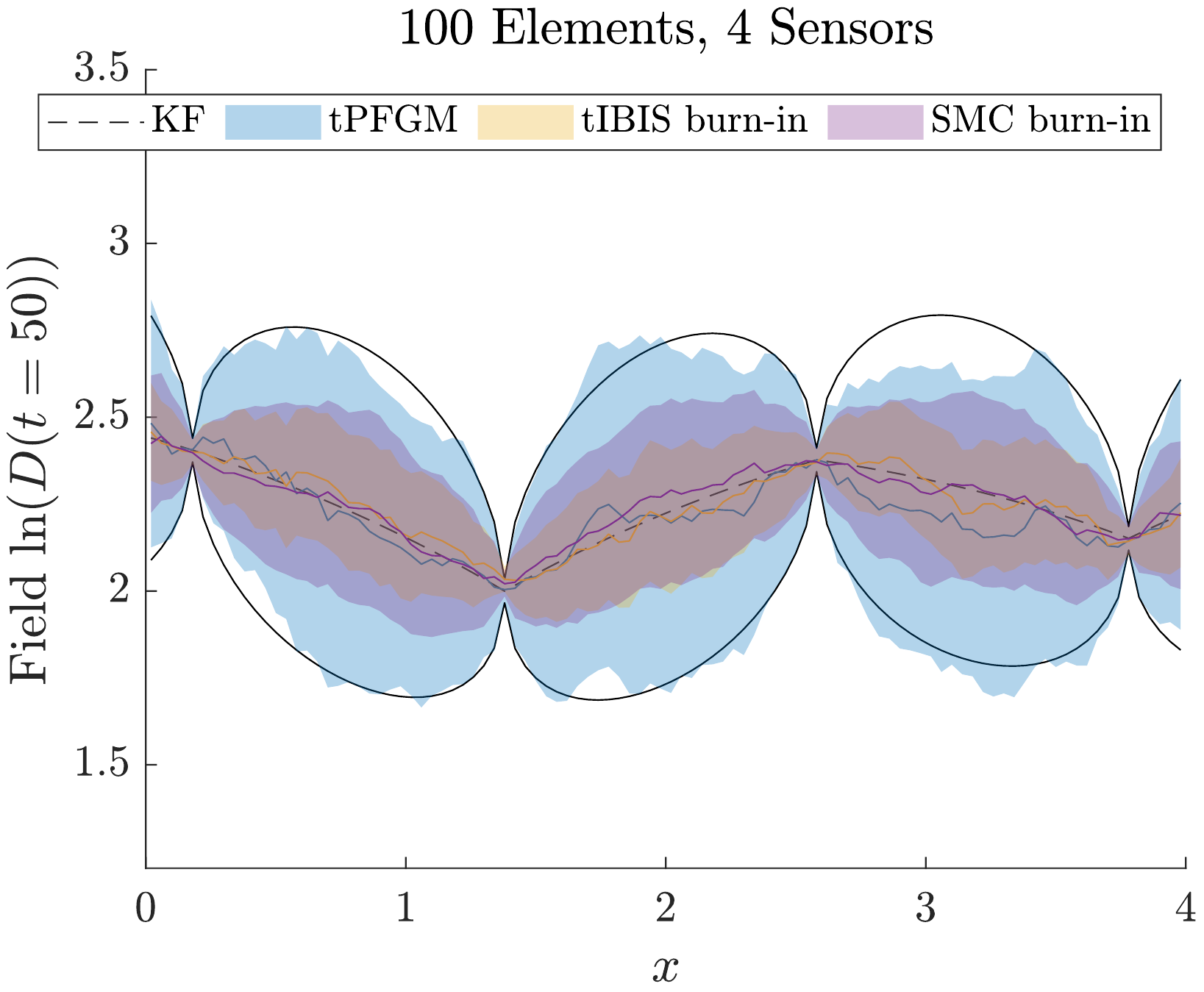}
	\end{subfigure}
	\caption{Updating of the random field $\ln \left(D(t=50)\right)$ in three different cases of varying problem dimensionality. The solid lines show the mean and the shaded areas the 90\% credible intervals inferred from 10 repeated runs of each filter. The black dashed line represented the posterior mean obtained via the KF, and the black solid lines the KF 90\% credible intervals}
	\label{fig:RF_updating}
\end{figure}

Based on the discussion of Section \ref{subsec:IBIS}, we introduce a burn-in period of $n_\text{B}$=5 in the IMH-GM kernel of Algorithm \ref{alg:mh_gm} and perform 50 new runs of the tIBIS and SMC filters. One can expect that inclusion of a burn-in is more likely to ensure sufficient exploration of the intermediate posterior distributions. However, at the same time the computational cost of tIBIS and SMC increases significantly, with a much larger number of required model evaluations than in Table \ref{tab:hd_model_eval}. In Figures \ref{fig:RF_comp_filters_mean_burn_in}, \ref{fig:RF_comp_filters_std_burn_in} we plot the mean and credible intervals for the relative errors in the estimation of the mean and standard deviation of the parameters. Comparing Figures \ref{fig:RF_comp_filters_mean} and \ref{fig:RF_comp_filters_mean_burn_in}, inclusion of burn-in is shown to lead to an improved performance of tIBIS and SMC in estimating the mean of the parameters in all cases. This improvement is more evident in the lower-dimensional case with 25 elements, and lessens as the problem dimension increases. Hence, with burn-in one observes a deterioration of the tIBIS and SMC filters' performance with increasing dimensionality. This point becomes more evident when looking at the relative errors of the estimated standard deviation in Figure \ref{fig:RF_comp_filters_std_burn_in}. With burn-in, the tIBIS and SMC filters provide better results than the tPFGM filter in estimating the standard deviation in the case of 25 elements, but perform progressively worse as the dimensionality increases, where they underestimate the KF reference standard deviation. This underestimation is clearly illustrated in Figure \ref{fig:RF_updating}. The reason for this behavior is the poor performance of the IMH-GM algorithm in high dimensions, which is numerically demonstrated in \cite{Papaioannou_2016}. We suspect that this behavior is related to the degeneracy of the acceptance probability of MH samplers in high dimensions, which has been extensively discussed in the literature for random walk samplers, e.g., in \cite{Gelman_1997, Au_2001, Katafygiotis_2008, Beskos_2009, Cotter_2013, Papaioannou_2015}. Single application of the IHM-GM kernel without burn-in yielded acceptance rates of around 50\% for all cases. With inclusion of burn-in, in higher dimensions, the acceptance rate in IMH-GM drops significantly in the later burn-in steps, leading to rejection of most proposed particles. To alleviate this issue, one could consider using the preconditioned Crank Nicolson (pCN) sampler to perform the move step within the IBIS and SMC filters, whose performance is shown to be independent of the dimension of the parameter space when the prior is Gaussian \citep{Cotter_2013}.

Increase of dimensionality does not seem to have any influence on the results obtained with the tPFGM filter. The illustrated efficacy of the tPFGM filter in estimating the time-invariant parameters in all considered cases of increasing dimensionality is related to the nature of the studied problem. The tPFGM filter relies entirely on the GMM approximation of the posterior distribution within its resampling process, in that it simply ``accepts" all the $N_{\text{par}}$ GMM-proposed particles, unlike the tIBIS and SMC filters, which contain the degenerating acceptance-rejection step within the IMH-GM move step. Clearly, the worse the GMM fit, the worse the expected performance of the tPFGM filter. The particular case investigated here has a Gaussian reference posterior solution, hence the GMM fitted by EM proves effective in approximating the posterior with a relatively small number of particles, even when going up to $d$=200 dimensions, thus leading to a good proposal distribution for sampling $N_{\text{par}}$ new particles in tPFGM. As reported in Table \ref{tab:hd_model_eval}, the tPFGM filter is associated with a significantly lower computational cost than its MCMC-based counterparts.

\vspace{-0.8cm}

\section{Concluding remarks}
\label{sec:Conclusions}
In this article, we present in full algorithmic detail three different on-line and off-line Bayesian filters, specifically tailored for the task of parameter estimation only of time-invariant deterioration model parameters in long-term monitoring settings. More specifically, these are an on-line particle filter with Gaussian mixture resampling (PFGM), an on-line iterated batch importance sampling (IBIS) filter, and an off-line sequential Monte Carlo (SMC) filter, which applies simulated annealing to sequentially arrive to a single posterior density of interest. The IBIS and SMC filters perform Markov Chain Monte Carlo (MCMC) move steps via application of an independent Metropolis Hastings kernel with a Gaussian mixture proposal distribution (IMH-GM) whenever degeneracy is identified. A simulated annealing process (tempering of the likelihood function) is further incorporated within the update step of the on-line PFGM and IBIS filters for cases when each new measurement is expected to have a strong information content; this leads to the presented tPFGM and tIBIS filters. The SMC filter can be employed only for off-line inference, while the PFGM, tPFGM, IBIS and tIBIS filters can perform both on-line and off-line inference tasks.

With the aid of two numerical examples, a rigorous comparative assessment of these algorithms for off-line and on-line Bayesian filtering of time-invariant deterioration model parameters is performed. In contrast to other works, the main focus here lies on the efficacy of the investigated Bayesian filters in quantifying the full posterior uncertainty of deterioration model parameters, as well as on the induced computational cost. 

For the first non-linear, non-Gaussian and low-dimensional case study, the IBIS and SMC filters, which both contain IMH-GM-based MCMC move steps, are shown to consistently outperform the purely on-line PFGM filter in estimating the parameters' reference posterior distributions. However, they induce a computational cost of at least an order of magnitude larger than the PFGM filter, when the same initial number of particles is used in all three filters. With similar computational cost, i.e., when increasing the number of particles in PFGM, it achieves enhanced posterior accuracy, comparable to the IBIS and SMC filters. 

For the second case study, involving a linear, Gaussian and high-dimensional model, the results vary with increasing problem dimensionality and number of sensors. The on-line tPFGM filter achieves a consistently satisfactory quality with increasing dimensionality, a behavior explained by the linear Gaussian nature of the problem, while a slight drop in the posterior quality is observed for increasing amount of sensor information. The tIBIS and SMC filters are shown to consistently outperform the tPFGM filter in lower dimensions, they however perform progressively worse in higher dimensions, a behavior likely explained by the degeneracy of the acceptance probability of MH samplers in high dimensions. The computational cost of the tIBIS and SMC filters is an order of magnitude larger than the tPFGM filter.

Some general conclusions drawn from the delivered comparative assessment are listed below.
\begin{itemize}[noitemsep,topsep=0pt]
	\itemsep0em 
	\item The IBIS (and its tIBIS variant) and SMC filters, which contain MCMC move steps, offer better approximations of the posterior mean of the model parameters than the purely on-line PFGM (and its tPFGM variant) filter with the same number of samples, as shown in both studied examples.
	\item The independent Metropolis Hastings (IMH)-based MCMC move step performed within the IBIS, tIBIS and SMC filters proves inadequate in properly exploring the posterior parameter space in high-dimensional problems.
	\item The purely on-line PFGM (and its tPFGM variant) filter is competitive with MCMC-based filters, especially for higher-dimensional well-behaved problems. 
\end{itemize}

Finally, to support the reader with the selection of the appropriate algorithm for a designated scenario, we provide Table \ref{table:guidelines}, which contains an assessment of the methods presented in this paper in function of problem characteristics.

\begin{table}[!ht]
	\caption{Set of suggestions on choice of the appropriate method in function of problem characteristics.}
	\small
	\centering
	\begin{tabular}{ccccc}
		\hline
		\multicolumn{2}{c}{\textbf{Criterion}}                                                                                                                                                                                                      & \textbf{\begin{tabular}[c]{@{}c@{}}PFGM \\ (tPFGM)\end{tabular}}                                                                          & \textbf{\begin{tabular}[c]{@{}c@{}}IMH-GM-based \\ IBIS (tIBIS)\end{tabular}}                                       & \multicolumn{1}{c}{\textbf{\begin{tabular}[c]{@{}c@{}}IMH-GM-based \\ SMC\end{tabular}}} \\ \hline
		\multicolumn{2}{c}{On-line inference}                                                                                                                                                                                                       & \checkmark                                                                                                                                 & $\circ$                                                                                                             & \multicolumn{1}{c}{$\times$}                                                             \\ \hline
		\multicolumn{2}{c}{Computational cost}                                                                                                                                                                                                      & $C_1$                                                                                                                                     & $C_2$                                                                                                               & \multicolumn{1}{c}{$C_3$}                                                                \\ \hline
		\multicolumn{1}{c}{\multirow{4}{*}{\rotatebox[origin=c]{90}{Applicability to different problems}}} & \multicolumn{1}{c}{\begin{tabular}[c]{@{}c@{}}Mean estimation \\ in low-dimensional, nonlinear,\\ non-Gaussian problems\end{tabular}}                           & $Q_3$                                                                                                                                     & $Q_4$                                                                                                               & \multicolumn{1}{c}{$Q_4$\vspace{0.2cm}}                                                                \\ 
		\multicolumn{1}{c}{}                                                     & \multicolumn{1}{c}{\begin{tabular}[c]{@{}c@{}}Uncertainty quantification\\ in low-dimensional, nonlinear, \\ non-Gaussian problems\end{tabular}}                & \multicolumn{1}{c}{$Q_3$}                                                                                                                                     & \multicolumn{1}{c}{$Q_4$}                                                                                                              & \multicolumn{1}{c}{$Q_4$\vspace{0.2cm}}                                                                \\ 
		\multicolumn{1}{c}{}                                                     & \multicolumn{1}{c}{\begin{tabular}[c]{@{}c@{}}Mean estimation\\ in high-dimensional, \\ well-behaved problems\end{tabular}}                                     & \multicolumn{1}{c}{$Q_2$}                                                                                                                                    & \multicolumn{1}{c}{$Q_3$}                                                                                                             & \multicolumn{1}{c}{$Q_4$\vspace{0.2cm}}                                                           \\ 
		\multicolumn{1}{c}{}                                                     & \multicolumn{1}{c}{\begin{tabular}[c]{@{}c@{}}Uncertainty quantification \\ in high-dimensional, \\ well-behaved problems\end{tabular}}                         & $Q_3$                                                                                                                                     & $Q_1$                                                                                                               & \multicolumn{1}{c}{$Q_1$}                                                                \\ \hline
		\multicolumn{2}{c}{\begin{tabular}[c]{@{}c@{}}Increasing sensor \\ information amount\end{tabular}}                                                                                                                                         & $\times$ (\checkmark)                                                                                                                      & $\times$ (\checkmark)                                                                                                & \multicolumn{1}{c}{\checkmark}                                                            \\ \hline
		\multicolumn{1}{l}{}                                                       & \multicolumn{1}{l}{\begin{tabular}[c]{@{}l@{}}$Q_1$: low quality\\ $Q_2$: moderate quality\\ $Q_3$: moderate to high quality\\ $Q_4$: high quality\end{tabular}} & \multicolumn{1}{l}{\begin{tabular}[c]{@{}l@{}}\checkmark: applicable\\ $\circ$: partly applicable\\ $\times$: not applicable\end{tabular}} & \multicolumn{1}{l}{\begin{tabular}[c]{@{}l@{}}$C_1$: moderate\\ $C_2$: moderate to high\\ $C_3$: high\end{tabular}} & \multicolumn{1}{l}{}                                                                     
	\end{tabular}
	\label{table:guidelines}
\end{table}

This paper does not investigate the performance of these filters when applied to high-dimensional and highly non-Gaussian problems. Such problems are bottlenecks for most existing filters and we expect the investigated filters to be confronted with difficulties in approximating the posterior distributions.
	
\section*{Funding Statement}
The work of A. Kamariotis and E. Chatzi has been carried out with the support of the Technical University of Munich - Institute for Advanced Study, Germany, funded by the German Excellence Initiative and the T{\"U}V S{\"U}D Foundation.

\section*{Competing Interests}
The authors declare none.

\section*{Data Availability Statement}
The data and code that support the findings of this study are openly available in \url{https://github.com/antoniskam/Offline_online_Bayes}.
	
	\bibliography{mybibfile}
	
\end{document}